\documentclass[sigconf]{acmart}

\usepackage[ruled, linesnumbered]{algorithm2e} 
\usepackage{amsfonts}                          
\usepackage{enumitem}                          
\usepackage{graphicx}                          
\usepackage{listings}                          
\usepackage{setspace}                          
\usepackage{subfigure}                         
\usepackage{xcolor}                            
\usepackage{xspace}                            
\usepackage{threeparttable}
\usepackage{booktabs}                          

\theoremstyle{definition}
\newtheorem{definition}{Definition}

\SetKwInOut{Input}{Input}
\SetKwInOut{Output}{Output}
\SetKw{And}{and}
\SetKwComment{Comment}{/* }{ */}

\newcommand{\sysname}{\mbox{\textsc{Nexus}}\xspace}
\newcommand{\parwoindent}[1]{\noindent\textbf{#1.}}
\newcommand{\entityTypeCache}{col\_entity\_type\xspace}

\definecolor{codegreen}{rgb}{0,0.6,0}
\definecolor{codegray}{rgb}{0.5,0.5,0.5}
\definecolor{codepurple}{rgb}{0.58,0,0.82}
\definecolor{backcolour}{rgb}{0.95,0.95,0.92}

\lstdefinestyle{mystyle}{
    backgroundcolor=\color{backcolour},   
    commentstyle=\color{codegreen},
    keywordstyle=\color{magenta},
    numberstyle=\tiny\color{codegray},
    stringstyle=\color{codepurple},
    basicstyle=\ttfamily\footnotesize,
    breakatwhitespace=true,
    breakindent=0pt,
    breaklines=true,                 
    captionpos=b,                    
    keepspaces=true,                 
    numbers=left,                    
    numbersep=5pt,                  
    showspaces=false,                
    showstringspaces=false,
    showtabs=false,                  
    tabsize=2,
    frame=single, 
    escapeinside={(*@}{@*)} 
}

\newcommand{\myitem}[0]{\scalebox{1.5}{\textbullet} \ }

\AtBeginDocument{%
  }

\setcopyright{none}
\begin{document}

\title{\sysname: Inferring Join Graphs from Metadata Alone via Iterative Low-Rank Matrix Completion}

\author{Tianji Cong}
\authornote{Work done while interning at Microsoft.}
\affiliation{%
  \institution{University of Michigan}
  \city{Ann Arbor}
  \state{Michigan}
  \country{USA}
}
\email{congtj@umich.edu}

\author{Yuanyuan Tian}
\affiliation{%
  \institution{Microsoft Gray Systems Lab}
  \city{Mountain View}
  \state{California}
  \country{USA}
}
\email{yuanyuantian@microsoft.com}

\author{Andreas Mueller}
\affiliation{%
  \institution{Microsoft Gray Systems Lab}
  \city{Mountain View}
  \state{California}
  \country{USA}
}
\email{Andreas.mueller.ml@gmail.com}

\author{Rathijit Sen}
\affiliation{%
  \institution{Microsoft Gray Systems Lab}
  \city{Redmond}
  \state{Washington}
  \country{USA}
}
\email{rathijit.sen@microsoft.com}

\author{Yeye He}
\affiliation{%
  \institution{Microsoft Research}
  \city{Redmond}
  \state{Washington}
  \country{USA}
}
\email{yeyehe@microsoft.com}

\author{Fotis Psallidas}
\affiliation{%
  \institution{Microsoft Gray Systems Lab}
  \city{New York}
  \state{New York}
  \country{USA}
}
\email{Fotis.Psallidas@microsoft.com}

\author{Shaleen Deep}
\affiliation{%
  \institution{Microsoft Gray Systems Lab}
  \city{Madison}
  \state{Wisconsin}
  \country{USA}
}
\email{shaleen.deep@microsoft.com}

\author{H. V. Jagadish}
\affiliation{%
  \institution{University of Michigan}
  \city{Ann Arbor}
  \state{Michigan}
  \country{USA}
}
\email{jag@umich.edu}

\renewcommand{\shortauthors}{Cong et al.}

\begin{abstract}
  Automatically inferring join relationships is a critical task for effective data discovery, integration, querying and reuse. However, accurately and efficiently identifying these relationships in large and complex schemas can be challenging, especially in enterprise settings where access to data values is constrained. In this paper, we introduce the problem of join graph inference when only metadata is available. We conduct an empirical study on a large number of real-world schemas and observe that join graphs when represented as adjacency matrices exhibit two key properties: high sparsity and low-rank structure. Based on these novel observations, we formulate join graph inference as a low-rank matrix completion problem and propose \sysname, an end-to-end solution using only metadata. To further enhance accuracy, we propose a novel Expectation-Maximization algorithm that alternates between low-rank matrix completion and refining join candidate probabilities by leveraging Large Language Models. Our extensive experiments demonstrate that \sysname outperforms existing methods by a significant margin on four datasets including a real-world production dataset. Additionally, \sysname can operate in a fast mode, providing comparable results with up to 6x speedup, offering a practical and efficient solution for real-world deployments.
\end{abstract}

\maketitle

\begin{figure*}[t]
\centering
\subfigure[0.33\textwidth][ER diagram]{
  \includegraphics[width=0.67\columnwidth]{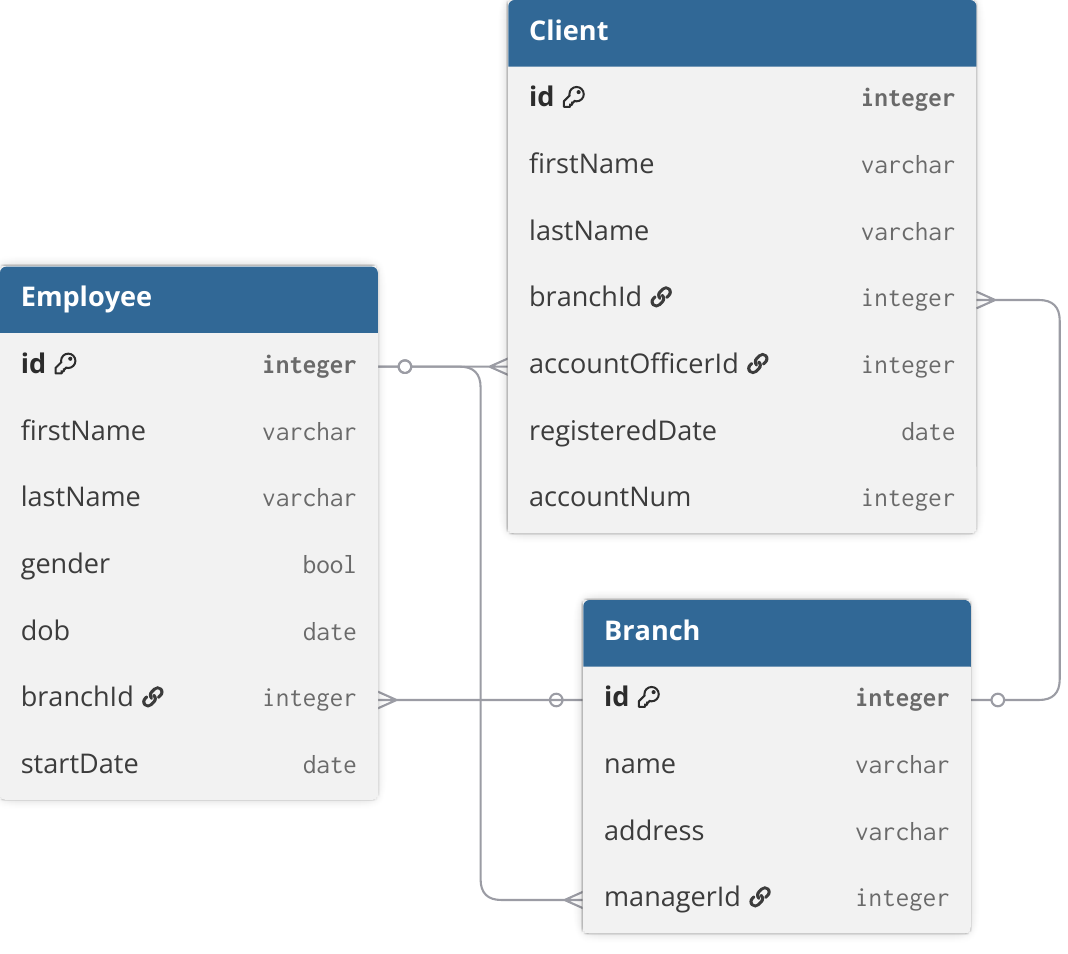}  
  \label{fig:er}
}\hfill
\subfigure[0.33\textwidth][Join graph]{
  \includegraphics[width=0.67\columnwidth]{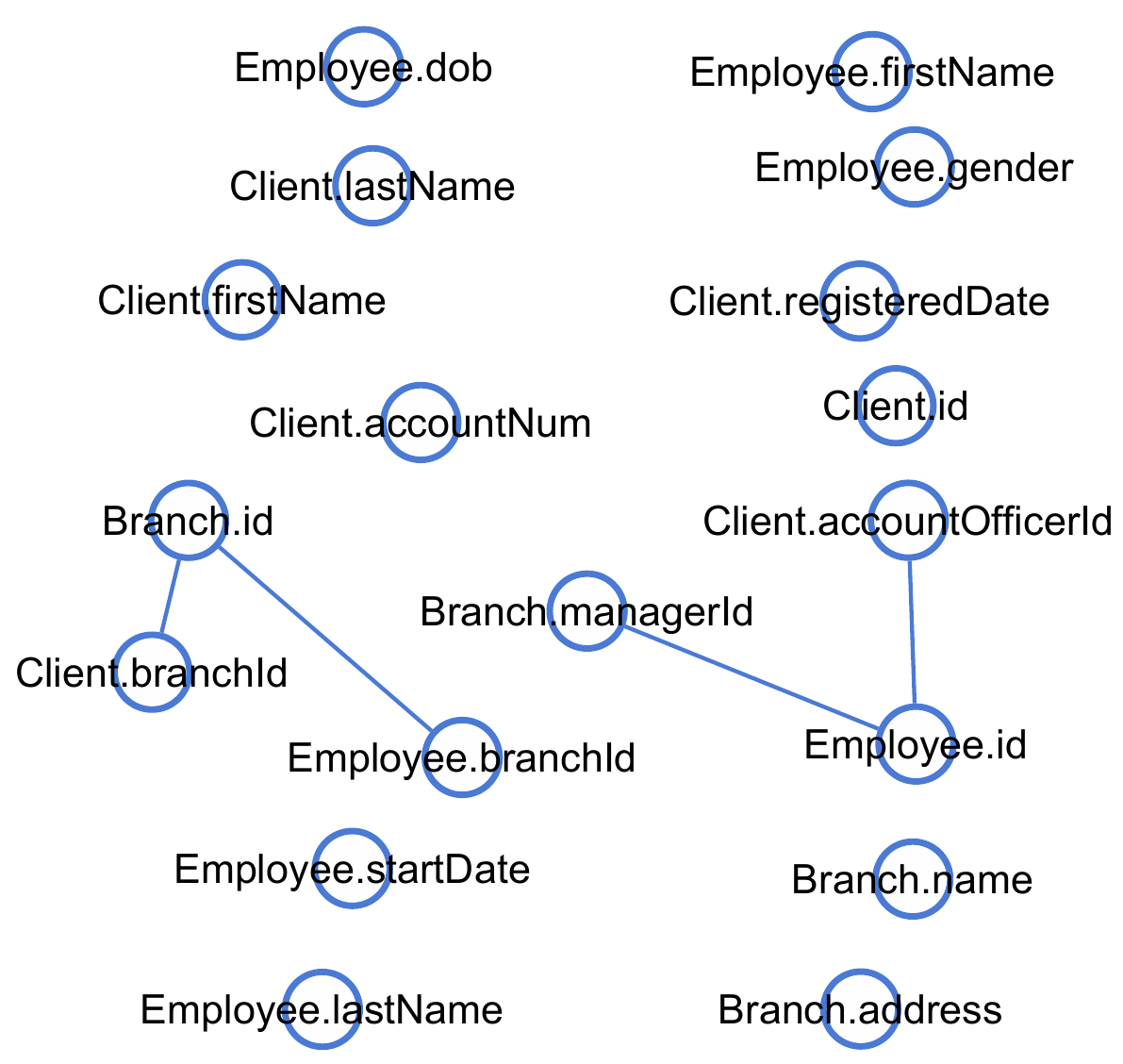} 
  \label{fig:joingraph}
}\hfill
\subfigure[0.33\textwidth][Join graph matrix (adjacency matrix)]{
  \includegraphics[width=0.67\columnwidth]{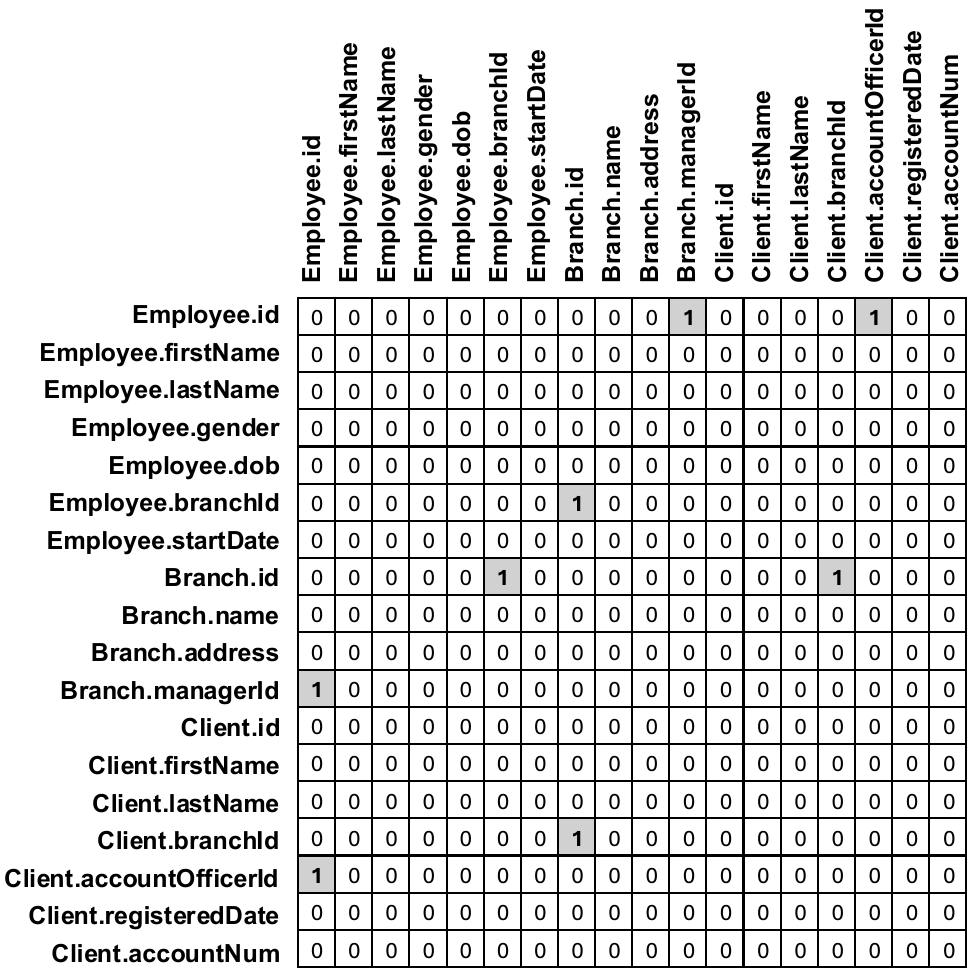} 
  \label{fig:adj_matrix}
}
\caption{An example schema of a bank database with its corresponding join graph and join graph matrix.}
\label{fig:example}
\end{figure*}

\section{INTRODUCTION}
  Detecting join relationships within a collection of tables is essential for data management tasks such as identifying primary and foreign keys (PK/FKs) in relational databases~\cite{rostin2009machine, zhang2010multi, chen2014fast, jiang2020holistic}, constructing Business Intelligence (BI) models for analytical workloads~\cite{bharadwaj2021discovering, lin2023autobi}, and finding related data in diverse sources like Web corpora~\cite{cafarella2008webtables, cafarella2009octopus, sarma2012finding, pimplikar2012answering} and data lakes~\cite{zhu2016lsh, fernandez2018aurum, zhu2019josie, bogatu2020d3l, zhang2020finding, dong2O21efficient, cong2023warpgate, dong2O23deepjoin, deng2024lakebench}.
  
  This capability is particularly important for modern lakehouse platforms. The very design principles that make the lakehouse architecture compelling (flexibility, scalability, and the unification of diverse data) create a challenging environment for join detection. Lakehouse operates on ``loose schema", where referential constraints like PK/FKs are often unsupported or undeclared. Furthermore, the consolidation of disparate datasets into a single repository obscures the formal boundaries between databases or schemas. Consequently, inferring the complete set of join relationships, which we term the \textit{join graph}, is a foundational capability in a lakehouse enabling automated data cataloging, enhanced data exploration, ML feature engineering, knowledge graph construction, and natural language querying. Figure~\ref{fig:joingraph} illustrates a join graph for a simple schema, where each node represents a table column and each edge indicates a join relationship between two columns.

  Join detection is not a new problem, and it has been tackled by a variety of methods, ranging from traditional heuristics~\cite{chen2014fast, jiang2020holistic, Microsoft/SemPy}, embedding-based techniques~\cite{fernandez2018aurum, dong2O23deepjoin, cong2023warpgate}, and ML models~\cite{rostin2009machine, bharadwaj2021discovering, maynou2024freyja} to, most recently, large language models (LLMs)~\cite{dohmen2024schemapile}. However, existing solutions predominantly rely on accessing data values. Utilizing data values---whether for direct value overlap calculation, embedding generation, or ML feature computation---is computationally expensive, especially for large datasets containing numerous and sizable tables. More importantly, the assumption that join detection algorithms can freely access data is often \textit{unrealistic}. For data service providers operating under strict privacy and security constraints, accessing customer data is often infeasible. Beyond data dependence, existing solutions typically perform only ``local'' pairwise detection, ignoring the global join graph structure. While Auto-BI~\cite{lin2023autobi} addresses this limitation through global graph-based optimization, it is purpose-built for BI settings where schemas follow strict arborescent structures (e.g., snowflake schemas). This design assumption breaks down in lakehouse contexts, where schemas exhibit significantly looser, less constrained structures.

  Unlike joinable table search~\cite{fernandez2018aurum, zhu2019josie, bogatu2020d3l, dong2O21efficient, flores2021nextiajd, cong2023warpgate, dong2O23deepjoin, deng2024lakebench, maynou2024freyja} which finds tables joinable with a given input table/column, we address a more challenging problem of \textit{join graph inference} in this paper. Given a collection of tables (specifically, their schemas and metadata), our objective is to infer the adjacency matrix of the corresponding join graph by uncovering equi-joins, particularly PK/FK relationships, among the tables. While these two problems are closely related, join graph inference is fundamentally harder: it requires discovering all pairwise join relationships across an entire table collection, resulting in quadratic worst-case computational complexity $O(n^2)$ with respect to the number of columns $n$, whereas joinable table search operates in linear worst-case time $O(n)$ by checking joinability against a specified input table/column. To this end, we develop \sysname, an \textit{accurate} and \textit{efficient} end-to-end join graph inference solution tailored for the enterprise setting. \sysname is designed to operate using \textit{only metadata} (detailed in Section~\ref{subsec:prob_statement}). Nevertheless, \sysname is flexible enough to incorporate \textit{query logs} and data values to further improve accuracy when they are available (Section~\ref{subsec:when_query_log_available} and~\ref{subsec:when_data_available}).

  The design of \sysname is guided by the crucial insight that \textit{real-world join graphs are highly structured}. Schemas are typically built around a small number of join keys, while most columns never participate in joins. This structure exhibits two key properties. First, the adjacency matrix of the join graph (or simply, the join graph matrix) is highly \textit{sparse}, as illustrated in Figure~\ref{fig:adj_matrix}. Second, besides those zero rows shown in Figure~\ref{fig:adj_matrix}, the matrix exhibits significant redundancy; for example, columns \texttt{Employee.branchId} and \texttt{Client.branchId} share the same joinable column set, making their corresponding rows in the matrix identical. This redundancy implies that the join graph structure can be captured by a small number of latent factors. Mathematically, this suggests the join graph matrix has a \textit{low rank} structure and can be approximated by a low-dimensional representation.

  Our extensive analysis of 5,957 non-trivial real-world database schemas from the SchemaPile-Perm dataset~\cite{dohmen2024schemapile} empirically confirms this insight: their join graphs are consistently sparse and low-rank. Around 98\% of the databases have a density below 0.02, and over 95\% have a normalized rank under 0.3. This study, detailed in Section~\ref{subsec:observation}, is to our knowledge the first and most extensive study of its kind and thus a significant contribution in itself. Establishing the high sparsity and low-rank nature of join graphs offers broader applicability than approaches restricted to strict arborescent structures common in BI settings~\cite{lin2023autobi}. In addition, our analysis informs our prioritization of detecting single-column joins. Among all databases containing join information in SchemaPile-Perm, only 7.5\% contain multi-column joins, representing just 3.5\% of all join cases. Consequently, while the current implementation of \sysname is optimized for this prevalent scenario, its design remains extensible to multi-column joins, as discussed in Section~\ref{subsec:multi_column_joins}.

  Based on our crucial insight, we formulate join graph inference as a \textit{low-rank matrix completion} problem, where the goal is to recover unknown entries of the join graph matrix from a set of known ones under a low-rank constraint. We initialize the matrix by assigning 0 to pruned column pairs and, where available, 1 to confirmed joins derived from query logs; while our method functions independently of query logs, they integrate naturally when present. For the remaining candidate entries that survive pruning, we utilize pretrained ML models or LLMs to score their likelihoods of being true joins. We then introduce a novel objective function that augments the standard low-rank completion goal with two additional terms: (1) a loss term that encourages the recovered entries to align with the probability scores given by ML models/LLMs, and (2) a regularization loss term that promotes the sparsity observed in real-world join graphs. This formulation allows us to effectively leverage predictions from pretrained models while adhering to the inherent structural properties of the join graph.

  In practice, pretrained models often struggle to generalize to unseen datasets, and forcing latent entries to conform to inaccurate predictions from these models can significantly degrade the quality of the inferred join graph. To address this challenge, we propose an Expectation-Maximization (EM) algorithm that iteratively refines the low-rank matrix completion by incorporating semantic guidance from LLMs. The EM algorithm alternates between two steps: (1) performing low-rank matrix completion, and (2) updating the prior probability estimates of join candidates using an LLM. A key component of this approach is leveraging an LLM to assess the semantic compatibility of \textit{column entity types}, which is a prerequisite for valid joins. We define a column entity type as a free-text annotation that encapsulates the semantic meaning of a column's values within the context of its table (e.g., annotating an \texttt{ID} column in an \texttt{Employee} table as ``employee ID''). Unlike existing annotation methods that depend on fixed sets of entity types or taxonomies, \sysname utilizes LLMs to handle diverse column semantics. We leverage LLMs to generate dynamic entity annotations and determine their semantic compatibility. Based on this assessment, we adjust the join probability scores---upweighting strong semantic matches and penalizing mismatches. This iterative refinement enables \sysname to correct inaccurate initial predictions and adapt to the unique semantics of unseen datasets.

  To summarize, this paper makes the following contributions:
  \begin{itemize}[left=10pt]
    \item We propose \sysname, an accurate and efficient end-to-end solution for join graph inference, tailored for privacy‑conscious enterprise settings. By relying on metadata only, and optionally leveraging query logs when available, \sysname operates effectively without direct access to data values.

    \item Our extensive empirical analysis of nearly 6K real-world database schemas confirms our crucial and novel insight: join graphs are both sparse and consistently low-rank. This new discovery forms the foundation for our novel join graph inference algorithm (Section~\ref{sec:prelim}).
    
    \item We formulate join graph inference as a low-rank matrix completion problem, with an optimization objective that explicitly exploits observed sparsity and low-rank structure while using pretrained models for initial join likelihoods (Section~\ref{sec:design}).
    
    \item To overcome the challenge of poor prior predictions from pretrained models, we propose a novel EM algorithm that alternates between performing low-rank matrix completion and leveraging an LLM to iteratively refine the probability scores of join candidates (Section~\ref{subsec:em_algo}).

    \item We demonstrate the effectiveness and efficiency of \sysname on four datasets including a real production dataset. Our metadata-only approach significantly outperforms baselines including those using data values (Section~\ref{sec:experiemnts}).
  \end{itemize}

\section{PRELIMINARIES AND OBSERVATIONS} \label{sec:prelim}
  This section establishes the foundation for our join graph inference solution. We will start by introducing key concepts and formally defining the problem. Following that, we will present findings from an empirical study on a large number of real-world database schemas, which directly motivate the design of our algorithm.

\subsection{Problem Statement}\label{subsec:prob_statement}
  We first give the definitions of join graphs, join graph matrices, and join graph probability matrices. Figure~\ref{fig:example} shows the join graph and join graph matrix for a simple example database schema.

  \begin{definition}[Join Graph]
    Given a set of tables $T$, a join graph is an undirected graph $G(T) = (V, E)$ where:

      \myitem $V = \{c_1, c_2, ..., c_n\}$ represents the set of all columns from the tables in $T$.
      
      \myitem $E$ is the set of join relationships. An edge $(c_i, c_j)$ exists between two columns $c_i$ and $c_j$, if and only if a join relationship exists between them.
  \end{definition}
  When the context is clear, we will refer to the join graph $G(T)$ simply as $G$.

  \begin{definition}[Join Graph Matrix]
    Let $G = (V, E)$ be the join graph for a set of tables $T$. The join graph matrix $A(T)\in \{0,1\} ^{n \times n}$, or simply denoted as $A$ when there is no ambiguity, is the adjacency matrix of the join graph $G$:
    \begin{equation}
      A_{i, j} = \begin{cases}
        1 \quad\quad \text{if $(c_i, c_j) \in E$} \\
        0 \quad\quad  \text{otherwise}
      \end{cases}  
    \end{equation}
  \end{definition}
  Since we consider the join graph undirected, the resulting matrix $A$ is symmetric ($A_{i,j} = A_{j,i}$) with a zero diagonal ($A_{i,i} = 0$).

  \begin{definition}[Join Graph Probability Matrix]
    Let $G = (V, E)$ be the join graph for a set of tables $T$. The join graph probability matrix $S(T)\in [0,1] ^{n \times n}$, or simply denoted as $S$ in case of no ambiguity, is an $n\times n$ matrix, where each entry $S_{i,j}$ denotes the probability that column $c_i$ and column $c_j$ are joinable. 
  \end{definition}
  $S$ is also a symmetric matrix ($S_{i,j} = S_{j,i}$) with a zero diagonal ($S_{i,i} = 0$) due to the undirected nature of join graphs.

  \begin{definition}[Metadata-Only Join Graph Inference]
    The problem of \textit{metadata-only join graph inference} is to reconstruct the join graph matrix $A$ for a given set of tables $T$, using \textit{only metadata} of $T$.
  \end{definition}

  The metadata we consider in this work includes table names, column names, column data types, and simple column statistics (total number of values, number of distinct values, number of nulls, and minimum and maximum values). These basic statistics are commonly available in the catalog service of a lakehouse platform for the purpose of query optimization. We simply reuse them for join graph inference. Critically, and in line with enterprise privacy and security requirements, our approach assumes no access to the actual data values within the columns.

  As noted before, our primary goal is to detect 1:1 and N:1 relationships, which correspond to PK/FK joins. Many-to-many (N:N) relationships can then be inferred via transitive closure. While we focus on single-column joins, the method is extensible to multi-column joins, as discussed in Section~\ref{subsec:multi_column_joins}.

\subsection{Empirical Analysis}\label{subsec:observation}
  \begin{table}[t]
    \centering
    \caption{Statistics of the curated database schemas.}
    \label{tab:schemapile_stats}
    \resizebox{0.95\columnwidth}{!}{%
      \begin{tabular}{lrrrr}
      \toprule
      \textbf{Statistics per database} & \textbf{Min} & \textbf{Max} & \textbf{Mean} & \textbf{Median} \\
      \midrule
      \textbf{\# tables} & 4 & 977 & 15 & 9 \\
      \textbf{\# columns} & 17 & 4572 & 108 & 56 \\
      \textbf{Avg. \# columns / table} & 4.04 & 37.46 & 6.73 & 6.00 \\
      \textbf{Density} & $1.56 \times 10^{-7}$ & 0.032 & 0.005 &  0.003 \\ 
      \textbf{Normalized rank} & $5.58 \times 10^{-4}$ & 0.49 & 0.12 & 0.10 \\
      \bottomrule
    \end{tabular}}
  \end{table}

  \begin{figure}[t]
    \centering
    \subfigure[0.5\columnwidth][CDF of Density]{
      \includegraphics[width=0.475\columnwidth]{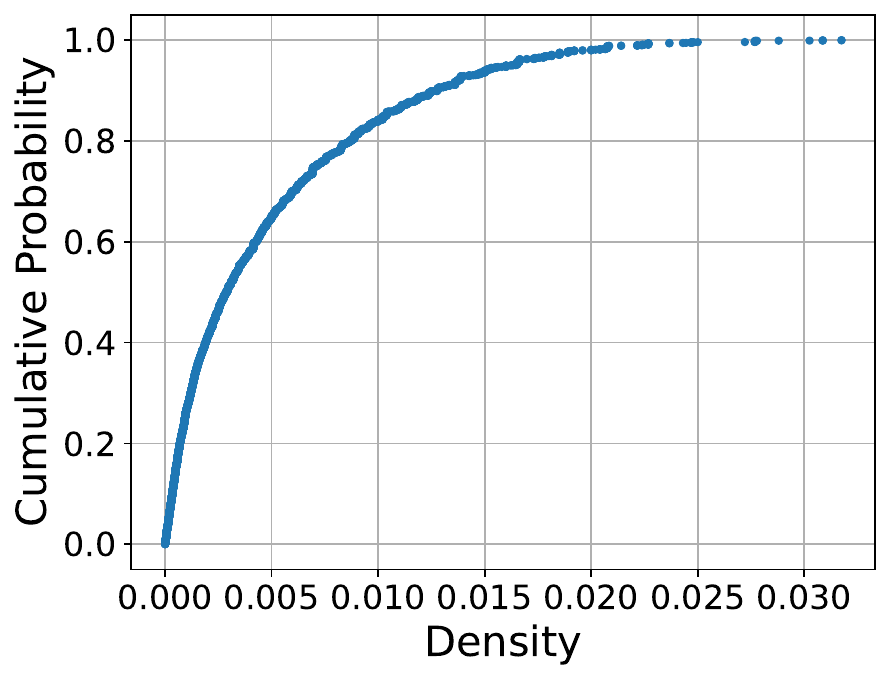}  
      \label{fig:sparsity}
    }\hfill
    \subfigure[0.5\columnwidth][CDF of Normalized Rank]{
      \includegraphics[width=0.475\columnwidth]{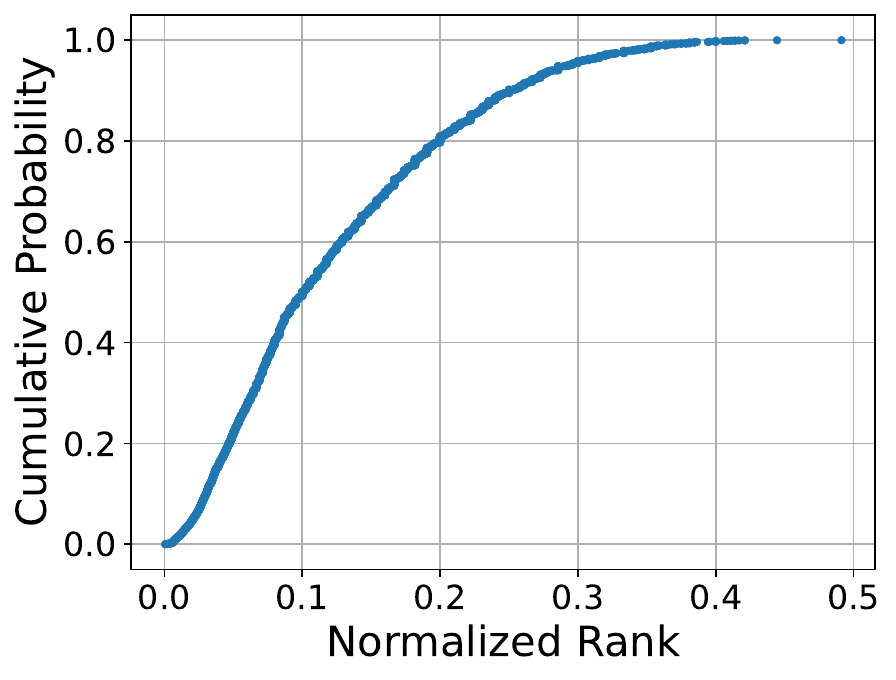} 
      \label{fig:rank}
    }
    \caption{CDF of density and normalized rank of the join graph matrices for the curated real-world database schemas.}
    \label{fig:sparsity-rank-analysis}
  \end{figure}

  To validate our hypothesis of join graph structure, we conducted an extensive empirical study to understand the characteristics of real-world join graphs. 

  For our empirical study, we curated a dataset of realistic, non-trivial database schemas from the SchemaPile-Perm collection~\cite{dohmen2024schemapile}. Starting with 22,989 databases, We first filtered this dataset to the 11,809 databases (around 51\%) that contain join relationships. We then focused on single-column joins by excluding the 7.5\% of schemas with multi-column relationships. Finally, we filtered out trivial databases, removing those with three or fewer tables or an average of four or fewer columns per table. This process resulted in a focused dataset of 5,957 databases for our analysis.

  As detailed in Table~\ref{tab:schemapile_stats}, this curated dataset represents a large and diverse collection of real-world database schemas: the number of tables per database ranges from 4 to 977, and the number of columns per database ranges from 17 to 4572. Besides basic statistics, our analysis focused on two key properties of join graph matrices: \textit{sparsity} and \textit{rank} (see Figure~\ref{fig:sparsity-rank-analysis}).

  \myitem \textit{High Sparsity:} We measured sparsity using matrix density, defined as $\frac{\text{\# nonzeros}}{n^2}$ for an $n \times n$ matrix. The results confirm that real-world join graphs are extremely sparse, with an average density below 0.005 and a median density under 0.003. As shown in Figure~\ref{fig:sparsity}, about 98\% of databases having a density less than 0.02.

  \myitem \textit{Low-Rank Structure:} We assessed the low-rankness of join graphs using the normalized rank, which is defined as $\frac{\text{rank}(A)}{n}$ for an $n \times n$ matrix $A$. This metric indicates how close a matrix is to being full-rank. We found that the average and median normalized ranks were only 0.12 and 0.10, respectively. As Figure~\ref{fig:rank} illustrates, over 95\% of the databases have a normalized rank below 0.3.

  These two key observations, high sparsity and low-rankness of join graphs, confirms our intuition that most columns in a schema are not joinable and that the overall join structure contains significant redundancy, allowing it to be represented in a low-dimensional space. In the following section, we describe our novel join graph inference algorithm based on these two key observations.

\begin{figure}[t]
  \centering
  \includegraphics[width=\columnwidth]{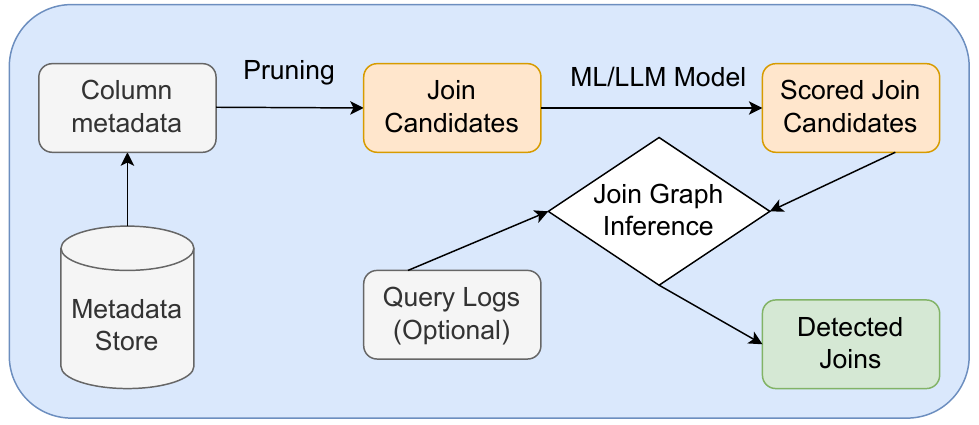}
  \caption{Overview of \sysname pipeline that infers join graphs from only metadata.}
  \label{fig:solution_overview}
\end{figure}

\section{\sysname Design}\label{sec:design}
  In this section, we provide an overview of our join graph inference solution, \sysname, and detail each core system component.

\subsection{Overview}
  Figure~\ref{fig:solution_overview} illustrates the overall workflow of \sysname. The process begins by retrieving schema information and column-level metadata for a given collection of tables from the metadata store. These  metadata are then used to prune unlikely join column pairs, significantly reducing the number of potential join candidates. Next, each candidate is scored using a pre-trained model, which can be either an ML model or an LLM. These scored candidates, optionally supplemented with known joins extracted from query logs, form a join graph probability matrix. In this matrix, non-candidate pairs are assigned a probability of 0, known joins from query logs are given a probability of 1.0, and the remaining entries are populated with scores from the pre-trained model. The probability matrix is subsequently fed into a matrix completion algorithm, which is based on our observations of high sparsity and low-rank structure within the join graph matrix. This algorithm is further enhanced by an iterative Expectation-Maximization (EM) procedure designed to mitigate any inaccuracies from the pre-trained model. The output is the final inferred join graph matrix, which identifies the detected join relationships.

\subsection{Metadata and Join Candidates Pruning}\label{subsec:join_candidates_pruning}
  Exhaustively evaluating every pair of columns as a potential join candidate incurs a quadratic cost relative to the number of columns, making it impractical for large datasets with thousands of columns. To address this, we collect column-level metadata from the metadata stores and apply cheap filters to eliminate a large number of unlikely join candidates early on. 
  
  We begin by discarding columns that are unlikely to serve as join keys, such as nested types (e.g., \texttt{JSON}, \texttt{ARRAY}, \texttt{MAP}), measure types (e.g., \texttt{FLOAT}, \texttt{DOUBLE}, \texttt{DECIMAL}), as well as \texttt{BLOB} and \texttt{BOOLEAN} types. For the remaining columns, we leverage the data type and the basic statistics from the metadata store, including cardinality (number of unique values), null value ratio, and min/max values. These statistics are typically maintained in metadata stores for query optimization purposes. Column pairs with incompatible data types (e.g., INTEGER vs. VARCHAR) are pruned immediately. Additionally, when inferring N:1 or 1:1 joins, we apply further pruning criteria, including ensuring uniqueness on the "1" side, verifying that the cardinality on the "N" side does not exceed that of the "1" side, and confirming that the value domain of the "N" side is a subset of the "1" side. These pruning criteria can be implemented in SQL syntax and executed efficiently in database systems.
  
  While more sophisticated profiling and pruning techniques exist~\cite{kruse2018pyro, jiang2020holistic, kaminsky2023sawfish}, they typically incur higher computational costs. We opt for these simpler, more efficient metadata-only filters because they effectively eliminate a large proportion of true negatives with minimal overhead.

\subsection{Scoring Join Candidates}
  The join candidates produced by the initial pruning step may still contain many false positives. For instance, a user ID column and a product ID column may share compatible data types and value ranges, yet they are not semantically joinable. Prior works~\cite{rostin2009machine, bharadwaj2021discovering, lin2023autobi} have typically employed ML models that combine multiple signals, such as embedding similarities of column names and Jaccard similarities of column values, to predict join relationships and estimate the associated probability scores. However, prior works rely on computing ML features that require access to the actual data values. Such computations are expensive and become prohibitive for large tables, often containing millions or billions of rows in enterprise environments. In contrast to most prior work, we assume no access to data values. We employ two metadata-based strategies to estimate the probability scores of join candidates:

  \myitem \textit{Pre-trained ML Model:} We pre-train an ML model exclusively on metadata-derived features. These include pairwise features derived from column names (max containment, Jaccard similarity, the ratio of edit distance to max length, Jaro-Winkler similarity, and the cosine similarity of weighted embeddings), and column-wise features (data type encodings, cardinality ratios, and null-value number ratios).
    
  \myitem \textit{LLM:} We can leverage an LLM as a pre-trained model, prompting it directly for both predictions and associated confidence levels, categorized as low, medium, and high. These qualitative confidence levels are then mapped to predefined probability scores (e.g., low $\rightarrow$ 0.1, medium $\rightarrow$ 0.6, and high $\rightarrow$ 0.9). This indirect elicitation of confidence is motivated by the known limitations of LLMs in generating reliable, well-calibrated confidence scores~\cite{mahaut2024factualconfidence, cong2024openforge, pawitan2024llmconfidence}.

  After this step, we construct a join graph probability matrix using the predicted scores from the pre-trained ML or LLM model, optionally augmented with confirmed join column pairs extracted from query logs, if available. 
  
  Despite the usefulness of pretrained models, they face two significant limitations: (1) their predictions are inherently local, considering only the features of individual column pairs and ignoring the global structure of the join graph, and (2) models trained on one dataset may not generalize effectively to others, and their confidence scores can be misleading. We address these critical challenges in Sections~\ref{subsec:lrmc_algo} and~\ref{subsec:em_algo}, respectively.

\subsection{Join Graph Inference through Low-Rank Matrix Completion}\label{subsec:lrmc_algo}
  After initial pruning and scoring, we obtain a join graph probability matrix, $S$. This matrix is partially observed: some entries are considered certain, with pruned candidates having value 0 and known joins from query logs having value 1, while the rest hold probability scores from a pre-trained model. We define $\Omega$ as the set of indices for the certain entries in $S$: $\Omega = \{(i, j)\,|\, (c_i, c_j)$ $\text{are either pruned out or from query logs}\}$.

  The objective of join graph inference is to recover a matrix $M\in [0,1]^{n \times n}$ that is a high-quality approximation of the unknown ground truth join graph matrix $A$. This recovery process only affects the uncertain entries of $S$. Drawing from our key observations that real-world join graphs are highly sparse and low-rank, we formulate this inference task as a low-rank matrix completion problem.

  The standard formulation of low-rank matrix completion seeks to minimize the rank of $M$ while preserving the observed values. Formally, the optimization objective can be written as:
  \begin{equation}\begin{aligned}\label{eq:standard_objective}
    &\text{minimize} \quad \text{rank}(M) \\
    &\text{subject to} \quad M_{i, j} = A_{i, j} \; \text{for} \; (i, j) \in \Omega 
  \end{aligned}\end{equation}
  where rank$(\cdot)$ gives the rank of a matrix.

  This is a non-convex optimization problem due to the rank function and is known to be NP-hard~\cite{chistov1984lrmc}. A well-established convex relaxation approximates the rank function with the nuclear norm of the matrix~\cite{candes2009lrmc}, resulting in the following optimization objective:
  \begin{equation}\begin{aligned}
    &\text{minimize} \quad ||M||_{*} = \sum_{k=1}^{n} \sigma_{k}(M) \\
    &\text{subject to} \quad M_{i, j} = A_{i, j} \; \text{for} \; (i, j) \in \Omega
  \end{aligned}\end{equation}
  where $\sigma_{k}(\cdot)$ denotes the $k$-th largest singular value of a matrix and the nuclear norm $||\cdot||_{*}$ is the sum of all singular values.

  While alternative approaches to low-rank matrix completion exist, such as matrix factorization~\cite{koren2009mftechniques, jain2013minimization}, they require pre-specifying the rank of the matrix to be recovered. Such prior knowledge of the matrix rank is not readily available in our setting. On the other hand, nuclear norm relaxation is generally considered less scalable than matrix factorization, as it needs to estimate all the parameters corresponding to the latent entries, which increase quadratically with respect to the matrix dimension. However, this concern is largely mitigated in our context due to a crucial observation: real-world join graphs are extremely sparse. Specifically, we observed that the detected join candidates (after pruning) constitute less than 5\% of the entries in the matrix we need to recover. Consequently, the number of parameters in our optimization is dramatically reduced, making the nuclear norm approach both feasible and efficient. This key insight not only addresses the scalability issue but also motivates us to explicitly incorporate a sparsity-inducing constraint into our recovery objective, ensuring the final solution reflects this fundamental property of join graphs.

  Additionally, we have access to the confidence scores for join candidates in the join graph probability matrix $S$, provided by the pre-trained model. These scores serve as prior information to guide the recovery of the latent entries in $M$.

  Let $\bar{\Omega}$ denote the complement of $\Omega$, which contains the set of positions for the uncertain entries in our probability matrix $S$. To integrate the observed sparsity and the prior beliefs, we propose the following optimization objective:
  \begin{equation}\begin{aligned}\label{eq:full_objective}
    &\text{minimize} \quad ||\text{P}_{\bar{\Omega}}(S - M)||^{2}_{F} \;+\; \lambda_{1} \; ||M||_{*} \;+\; \lambda_{2} \; ||M||_{1} \\
    &\text{subject to} \quad M_{i, j} = A_{i, j} \; \text{for} \; (i, j) \in \Omega 
  \end{aligned}\end{equation}

  Here, $\text{P}_{\bar{\Omega}}$ is a projection operator that preserves the entries with positions in $\bar{\Omega}$ while zeroing out all others. $||\cdot||_{F}$ and $||\cdot||_{1}$ denote the Frobenius norm and the 1-norm of a matrix, respectively. 

  The above object function can be broken down into the following three components:
    \myitem \textit{Confidence Score Regularization}: The term $||\text{P}_{\bar{\Omega}}(S - M)||^{2}_{F}$ encourages the final matrix M to align with the confidence scores from the pre-trained model, but only for the uncertain entries in $\bar{\Omega}$.
    
    \myitem \textit{Low-Rank Regularization}: The term $\lambda_{1} \; ||M||_{*}$ uses the nuclear norm  to encourages the solution $M$ to be low-rank, aligning with our key observation.
    
    \myitem \textit{Sparsity Regularization}: The term $\lambda_{2} \; ||M||_{1}$ uses the L1 norm to promote sparsity in $M$, effectively pushing the probabilities of unlikely joins towards zero.

  $\lambda_{1}$ and $\lambda_{2}$ are hyperparameters weighting the regularization power of the low-rank term and the sparsity term, respectively.

  Note that the recovered matrix $M$ contains real values in $[0, 1]$, rather than binary indicators. To produce the final binary decisions, we apply a threshold $\theta$ to the latent entries of $M$. The resulting decision matrix $\hat{M} \in \{0,1\} ^{n \times n}$ is defined as:
  \begin{equation}
    \hat{M}_{i, j} = \begin{cases}
      M_{i, j} \quad\quad\quad\quad \text{if $(i, j) \in \Omega$} \\
      \mathbf{1}(M_{i, j} \geq \theta) \quad \text{otherwise}
    \end{cases}  
  \end{equation}
  where $\mathbf{1}(\cdot)$ is the indicator function.

  \begin{algorithm}[t]
    \setstretch{1.1}
    \caption{Expectation-Maximization Algorithm for Iterative Join Graph Inference}\label{alg:em_optimization}
    \Input{$S^{(0)}$, initial probability matrix;\\ $\Omega^{(0)}$, initial set of known certain entries in $S^{(0)}$;\\ $\Gamma$, max number of iterations;\\ $\epsilon$, tolerance for early stopping}
    \Output{$M$, recovered matrix}

    \BlankLine
    \tcc{Map from column to its entity type (cache)}
    $\entityTypeCache = \{\}$\;
    
    \For{$t = 0, 1, \dots, \Gamma-1$}{
      \BlankLine
      \tcc{E-step: Recover the latent matrix $M^{(t+1)}$}
      $M^{(t+1)}$ = \textbf{low\_rank\_matrix\_completion}($S^{(t)}, \Omega^{(t)}$)\;

      \BlankLine
      \tcc{Check for convergence}
      \If{$t > 0$}{
        \tcc{No candidates left or last iteration}
        \If{$\bar{\Omega}^{(t)} == \emptyset$ \textup{\textbf{or}} $t == \Gamma-1$}{
          \textbf{break}\;
        }
        \If{$\| M^{(t+1)} - M^{(t)} \|_{F} \,\leq\, \epsilon$}{
          \textbf{break}\;
        }
      }

      \BlankLine
      \tcc{M-step: Update the probability matrix}
      $S^{(t+1)}, \Omega^{(t+1)}, \entityTypeCache = \textbf{update\_prob\_matrix}(M^{(t+1)}, \Omega^{(t)}, \entityTypeCache)$
    }

    \BlankLine
    \Return{$M^{(t+1)}$}
  \end{algorithm}

  \begin{algorithm}[t]
    \setstretch{1.1}
    \caption{Update the Probability Matrix (\textbf{update\_prob\_matrix})}\label{alg:update_prob_matrix}
    \Input{$S$, the probability matrix;\\
    $\Omega$, the set of known certain entries in $S$;\\
    $\entityTypeCache$, column to entity type map;\\
    $low\_threshold$, the threshold to prompt LLM for\\ \ \ \ \ \ \ \ entity types and type match check;\\
    $high\_threshold$, the threshold to promote a latent\\ \ \ \ \ \ \ \  entry to be a hard positive;\\
    $\delta$, the penalty factor for entity type mismatch}
    \Output{updated $S$, $\Omega$, and $\entityTypeCache$}

    \BlankLine
    $join\_candidates = \{(i, j)\;|\; (i, j) \in \bar{\Omega}\; \text{and}\; i < j\}$\;

    \BlankLine
    \tcc{Loop through join candidates}
    \For{$(i, j) \in join\_candidates$}{
        \If{$S[i][j] < low\_threshold$}{
            \textbf{continue}\;
        }

        \If{$i \notin $ \entityTypeCache}{
            $\entityTypeCache[i] = \textbf{prompt\_llm\_for\_entity\_type}(col\_i)$\;
        }
        
        \If{$j \notin$ \entityTypeCache}{
            $\entityTypeCache[j] = \textbf{prompt\_llm\_for\_entity\_type}(col\_j)$\;
        }

        \tcc{Boost probability of type-matched entry}
        \eIf{$\entityTypeCache[i] == \entityTypeCache[j]$ \textbf{\textup{or}} \textbf{\textup{prompt\_llm\_for\_soft\_type\_match}}$(col\_i, col\_j)$}{  
            \eIf{$S[i][j] \geq high\_threshold$}{
                $S[i][j] = S[j][i] = 1.0$\;
                \tcc{Add new known entries}
                $\Omega.add(\{(i,j),\; (j,i)\} )$\;
            }{
                $S[i][j] = S[j][i] = high\_threshold$\;
            }
        }{
        \tcc{Decay probability of unmatched entry}
            $S[i][j] \mathrel{*}= \delta$; $S[j][i] \mathrel{*}= \delta$\;
        }
    }
    
    \Return{$S, \Omega, \entityTypeCache$}
  \end{algorithm}

\subsection{Iterative Join Graph Inference via Expectation-Maximization Algorithm}\label{subsec:em_algo}
  As shown in Equation~\ref{eq:full_objective}, we (partially) align the recovered matrix with the probability matrix determined by a pre-trained ML/LLM model. This approach assumes that the model accurately estimates the probability scores of join candidates. However, this assumption may not hold when applying a pretrained model to unseen datasets with very different distributions from the training datasets. Consequently, forcing latent entries in the recovered matrix to match potentially inaccurate estimates can degrade result quality. To address this issue, we propose an Expectation-Maximization (EM) algorithm (Algorithm~\ref{alg:em_optimization}), which iteratively improves the quality of low-rank matrix completion using guidance from an LLM.

  EM is a widely adopted approach in ML for computing maximum likelihood estimates in models with incomplete or latent variables~\cite{dempster1977maximum}. The algorithm, outlined in Algorithm~\ref{alg:em_optimization}, operates iteratively in two main steps. The \textit{Expectation step} (E-step) involves low-rank matrix completion to estimate the values of latent entries. This estimation is based on the observed data $\Omega$ and the probability scores $S^{(t)}$ given by an ML/LLM model. This E-step can be interpreted as determining the posterior probabilities of the latent entries given the observed data and the prior knowledge from the ML/LLM model. Subsequently, in the \textit{Maximization step} (M-step), we refine the probability matrix by incorporating a semantic compatibility assessment of column entity types using an LLM. Specifically, we harness the semantic power of LLMs to enforce a key condition for valid joins: \textit{two columns that can be joined must have semantically equivalent entity types}. 
  
  We define a \textit{column entity type} as a free-text annotation that captures the specific semantics which a column represents within its relational table. For instance, an ID column in a table of food orders should ideally be annotated as ``food order ID'', whereas an ID column in a table of users should be annotated as ``user ID''. Leveraging the advanced semantic understanding capabilities of LLMs, we employ an LLM to annotate column entity types and to determine the semantic equivalence between two entity types. The match of entity types constitutes a necessary condition for valid joins and substantially reduces the number of spurious join candidates (e.g., two ID columns have coincidental value overlap but refer to distinct types of entities). Critically, our approach does not rely on a fixed vocabulary of entity types, as column semantics can vary across different tables and databases, and no existing entity type vocabularies have a comprehensive coverage of the nuanced spectrum of column semantics.
  
  As detailed in Algorithm~\ref{alg:update_prob_matrix}, the M-step iterates through latent entries (join candidates) and prompts an LLM to identify the entity types of columns based on column and corresponding table names. For each candidate pair with a score exceeding a predefined \textit{low threshold}, we prompt an LLM to infer their entity types if they do not have one. To reduce the computational and monetary cost of LLM calls, we cache the identified entity types. Subsequently, we conduct a character-wise comparison of their entity types. If this comparison fails, we further prompt the LLM for a soft check of whether the entity types match semantically. The alignment of entity types is a prerequisite for a valid join. We have observed LLMs to be particularly effective in performing such nuanced semantic comparisons. If the entity types align and the associated probability score exceeds a predefined \textit{high threshold}, we promote the pair to be a hard positive, setting its probability score to 1 and removing it from the set of latent entries. If the types match but the probability score falls below the high threshold, we boost the probability score to this threshold value, ensuring the pair undergoes another LLM check in the next EM iteration. This design accounts for the low initial probability estimate of the pair from the pre-trained model and the inherent stochasticity of LLM outputs. If the entity type check fails, we penalize the pair by reducing its probability score by a specific factor $\delta$, which we set to 0.5 for all experiments. In the implementation, to further reduce the latency of LLM calls, we batch multiple requests in one prompt with a batch size of 24. 

  The iterative process of the EM algorithm continues until either a predefined maximum number of iterations, $\Gamma$, is reached or convergence is achieved. The convergence is determined by the change in the recovered matrix (quantified by the Frobenius norm) falling below a specified tolerance, $\epsilon$. We set $\epsilon$ to $1e^{-5}$ in all experiments.
  
  For latency-sensitive scenarios, our system can be configured to perform a single iteration of the EM algorithm (only low-rank matrix completion). This fast mode, which we call \sysname-Fast, is particularly useful when pre-trained models provide strong priors. It bypasses all LLM calls in the EM algorithm, ensuring maximum inference speed while still improving over the pretrained models and achieving competitive results compared to our full EM algorithm. In the following text, we refer to our full EM algorithm as \sysname-EM to distinguish from \sysname-Fast.

\subsection{Optimization via Core Submatrix Reduction}\label{subsec:fast_mode}
  The procedure described above recovers a matrix $M$ of size $n\times n$, where $n$ is the total number of columns in the table collection. For a large table collection, $n$ can be very big. However, since our primary goal is to infer latent entries determined by join candidates that typically involve only a small subset of all columns, we can instead recover a smaller matrix, which we call the \textit{core submatrix}:

  \begin{definition}[Core Submatrix]
    Given a symmetric $n \times n$ matrix $M$, the core submatrix $M'$ is defined as the submatrix obtained by removing all rows and columns from $M$ that contain only zero entries. Formally, let $I = \{ i \mid \exists j, M_{i, j} \neq 0 \}$ be the set of indices corresponding to non-zero rows (and equivalently columns, since $M$ is symmetric). Then, the core submatrix can be denoted as $M' = M[I, I]$. The dimension of $M'$ is $n'\times n'$, where $n' =|I|$.
  \end{definition}

  Since most columns are not join keys, especially those measure-oriented columns ubiquitous in analytical workloads, typically the core submatrix $M'$ is much smaller than $M$ (i.e. $n' << n$). Importantly, our approach based on low-rank matrix completion remains applicable, as we found that $M'$ also exhibits high sparsity and a low-rank structure akin to the full adjacency matrix.

  Recovering a smaller core submatrix $M'$ offers significant practical benefits. Empirically, its dimension is often less than half the size of $M$, resulting in faster runtime and reduced memory usage. The main caveat is that the component of pruning join candidates is not perfect, and some relevant join keys may be excluded from $M'$ and not recovered. Through experiments, we observe that this optimization gives almost 25\% speedup when only running low-rank matrix completion (i.e., Nexus-Fast), so we enable this optimization by default for Nexus-Fast. For \sysname-EM, this optimization is less significant as the main performance bottleneck lies in LLM calls (from a remote service). Hence, we choose to recover the full matrix without losing any relevant join keys.

\subsection{Extension to Multi-Column Joins}\label{subsec:multi_column_joins}
  While we have discussed our approach in the context of single-column joins, it's easily adaptable for multi-column join scenarios. The core idea remains the same: we treat every set of k columns within a table as a single conceptual unit. These units are then subject to the same pruning rules outlined in Section~\ref{subsec:join_candidates_pruning}. This extension leads to a larger probability matrix for the low-rank matrix completion task. Specifically, for a set of tables $T$, the matrix dimension becomes $m\times m$, where $m=\sum_{\tau\in T}{\binom{n_\tau}{k}}$ and $n_\tau$ is the number of columns in table $\tau$. Each entry in this matrix represents the join probability of a pair of $k$-column units.
  
  Crucially, this expanded matrix also exhibits high sparsity and a low-rank structure. This allows us to directly apply the same matrix completion and EM algorithms (Section~\ref{subsec:lrmc_algo} and Section~\ref{subsec:em_algo}) to detect multi-column joins. In fact, the presence of multiple columns in a unit increases the matrix size while the overall number of multi-column joins is typically smaller. This often leads to an even higher sparsity and lower rank in the join graph matrix, which can benefit our algorithms. Additionally, the core submatrix reduction optimization from Section~\ref{subsec:fast_mode} is also applicable and can be employed to further improve the efficiency.

  Despite this capability, our analysis of the SchemaPile-Perm dataset shows that multi-column joins constitute only a small fraction (7.5\%) of real-world schemas. Consequently, our experimental evaluation focuses on the more prevalent single-column joins.

\begin{table}[t!]
  \renewcommand{\arraystretch}{1.2}
  \centering
  \caption{Experimental datasets and their statistics.}
  \label{tab:dataset_stats}
  \resizebox{\columnwidth}{!}{%
    \begin{tabular}{crrrrrr}
    \toprule
    Datasets           & \# Schema & \# Tables & \# Columns & \# Candidates & Density & Normalized Rank \\ \midrule 
    TPC-H              & 1         & 8         & 61    & 1,830    & 0.004     & 0.20 \\
    TPC-DS             & 1         & 24        & 425   & 90,100     & 0.001   &  0.08 \\
    BIRD-SQL           & 11        & 75        & 806    & 324,415    & 0.003   &  0.10 \\
    REAL               & 3         & 30        & 1156   & 667,590    & 0.001     & 0.02 \\ \bottomrule
    \end{tabular}
  }
  \vspace{-3mm}
\end{table}

\begin{table*}[t]
\renewcommand{\arraystretch}{1.2}
\centering
\caption{Quality comparison of metadata-only approaches on four datasets.}
\label{tab:quality_comparison}
\resizebox{0.95\textwidth}{!}{%
\begin{tabular}{lllllllllllll}
\hline
  & \multicolumn{3}{c}{TPC-H} & \multicolumn{3}{c}{TPC-DS} & \multicolumn{3}{c}{BIRD-SQL} & \multicolumn{3}{c}{REAL} \\ \hline
  & \multicolumn{1}{c}{F1} & \multicolumn{1}{c}{Precision} & \multicolumn{1}{c|}{Recall} & \multicolumn{1}{c}{F1} & \multicolumn{1}{c}{Precision} & \multicolumn{1}{c|}{Recall} & \multicolumn{1}{c}{F1} & \multicolumn{1}{c}{Precision} & \multicolumn{1}{c|}{Recall} & F1 & Precision & \multicolumn{1}{l}{Recall} \\ \hline
  \multicolumn{1}{l|}{XGBoost} & 0.35 & 0.21 & \multicolumn{1}{l|}{1.0} & 0.44 & 0.34 & \multicolumn{1}{l|}{0.62} & 0.10 & 0.05 & \multicolumn{1}{l|}{0.99} & 0.29 & 0.17 & 0.89 \\
  \multicolumn{1}{l|}{GPT-4o} & 0.78 & 0.64 & \multicolumn{1}{l|}{1.0} & 0.81 & 0.68 & \multicolumn{1}{l|}{0.99} & 0.56 & 0.39 & \multicolumn{1}{l|}{0.98} & 0.49 & 0.34 & 0.88 \\
  \multicolumn{1}{l|}{SemPy} & 0.41 & 0.26 & \multicolumn{1}{l|}{1.0} & 0.43 & 0.50 & \multicolumn{1}{l|}{0.38} & 0.29 & 0.22 & \multicolumn{1}{l|}{0.47} & 0.26 & 0.29 & 0.23 \\
  \multicolumn{1}{l|}{Auto-BI} & 0.56 & 0.39 & \multicolumn{1}{l|}{1.0} & 0.52 & 0.61 & \multicolumn{1}{l|}{0.45} & 0.32 & 0.42 & \multicolumn{1}{l|}{0.26} & 0.08 & 0.15 & 0.05 \\
  \multicolumn{1}{l|}{Nexus-XGBoost (Ours)} & \textbf{0.88} & 0.78 & \multicolumn{1}{l|}{1.0} & 0.72 & 0.58 & \multicolumn{1}{l|}{0.95} & 0.57 & 0.44 & \multicolumn{1}{l|}{0.85} & 0.50 & 0.42 & 0.63 \\
  \multicolumn{1}{l|}{Nexus-GPT-4o (Ours)} & \textbf{0.88} & 0.78 & \multicolumn{1}{l|}{1.0} & \textbf{0.89} & 0.88 & \multicolumn{1}{l|}{0.90} & \textbf{0.83} & 0.86 & \multicolumn{1}{l|}{0.80} & \textbf{0.61} & 0.49 & 0.80 \\ \hline
\end{tabular}%
}
\end{table*}

\section{EXPERIMENTS} \label{sec:experiemnts}
  This section presents our experimental results, covering quality and efficiency comparisons as well as a hyperparameter ablation study. Unless otherwise specified (as in Section~\ref{subsec:when_data_available}), all experiments use the metadata-only setup.

\subsection{Experiment Setup}
  \parwoindent{Datasets} Table~\ref{tab:dataset_stats} lists the datasets used to evaluate \sysname, covering both standard benchmarks and complex multi-source scenarios. We include TPC-H and TPC-DS, following the state-of-the-art (SOTA) work~\cite{lin2023autobi}. Since these benchmarks represent single-schema environments, we also created a multi-source scenario by combining 11 databases from the development set of BIRD-SQL~\cite{li2024birdsql}. Additionally, we introduce REAL, a production dataset spanning three distinct sources where only metadata and query logs are available. For ground truth, we rely on defined PK/FK constraints for the synthetic datasets and derive relationships from query logs for REAL, as it lacks explicit constraints. REAL reflects the "messiness" of production settings, providing a rigorous test for join detection. Although we use the Schemapile-Perm dataset to analyze join graph structure, we cannot use it for evaluation because it lacks the actual data values and column statistics. Instead, we utilize this dataset for ML model training (see below). Finally, column 5 of Table~\ref{tab:dataset_stats} reports the worst-case number of candidate pairs, reflecting the growth in computational complexity as the dataset size increases.

  \parwoindent{Baselines} First, we evaluate standalone ML models: XGBoost and GPT-4o. For XGBoost, we trained on a large corpus of schemas from SchemaPile-Perm (Section~\ref{subsec:observation}), utilizing a superset\footnote{The features include containment/Jaccard similarity/edit distance ratio/jaro-winkler similarity of concatenated table and column name tokens, boolean indicators for column data types (integer and string), in addition to those used in~\cite{bharadwaj2021discovering}.} of metadata features from~\cite{bharadwaj2021discovering}. For GPT-4o, we used zero-shot prompts with metadata. These two models also serve as priors for our system (denoted \sysname-XGBoost and \sysname-GPT-4o).
  
  Second, we compare against SemPy~\cite{Microsoft/SemPy} (from Microsoft Fabric) and Auto-BI (the current SOTA). In addition to their original versions that utilize data values, we adapted them for our primary metadata-only evaluation. For SemPy, we disabled value-based overlap checks, relying on column name similarity (Ratcliff/Obershelp) and other heuristics. For Auto-BI, which frames join detection as a graph optimization problem, we replaced its original XGBoost model with our XGBoost model trained on metadata features.
  
  Finally, we include Aurum~\cite{fernandez2018aurum} and DeepJoin~\cite{dong2O23deepjoin}. These methods focus on the related ``joinable table search'' problem and primarily rely on data values; therefore, we restrict their evaluation to the experiments where data access is permitted (Section~\ref{subsec:when_data_available}). We also attempted to evaluate JOSIE~\cite{zhu2019josie} but excluded it as it consistently timed out (>1 hour for multi-source datasets), a limitation also noted in prior work~\cite{deng2024lakebench}.

  \parwoindent{Metrics} Due to the class imbalance in detected join candidates, we evaluate the quality of all approaches using F1 score, precision, and recall. The reported numbers represent the best results for each method after we performed best-effort tuning on each dataset. For efficiency comparison, we report runtime in seconds.

  \parwoindent{Hardware} All experiments were conducted on an Azure virtual machine (standard NC12s v3) with 12 vCPUs and 224 GiB of RAM.

  \begin{figure*}[ht]
    \begin{minipage}[t]{0.245\textwidth}
      \centering
      \includegraphics[width=\columnwidth]{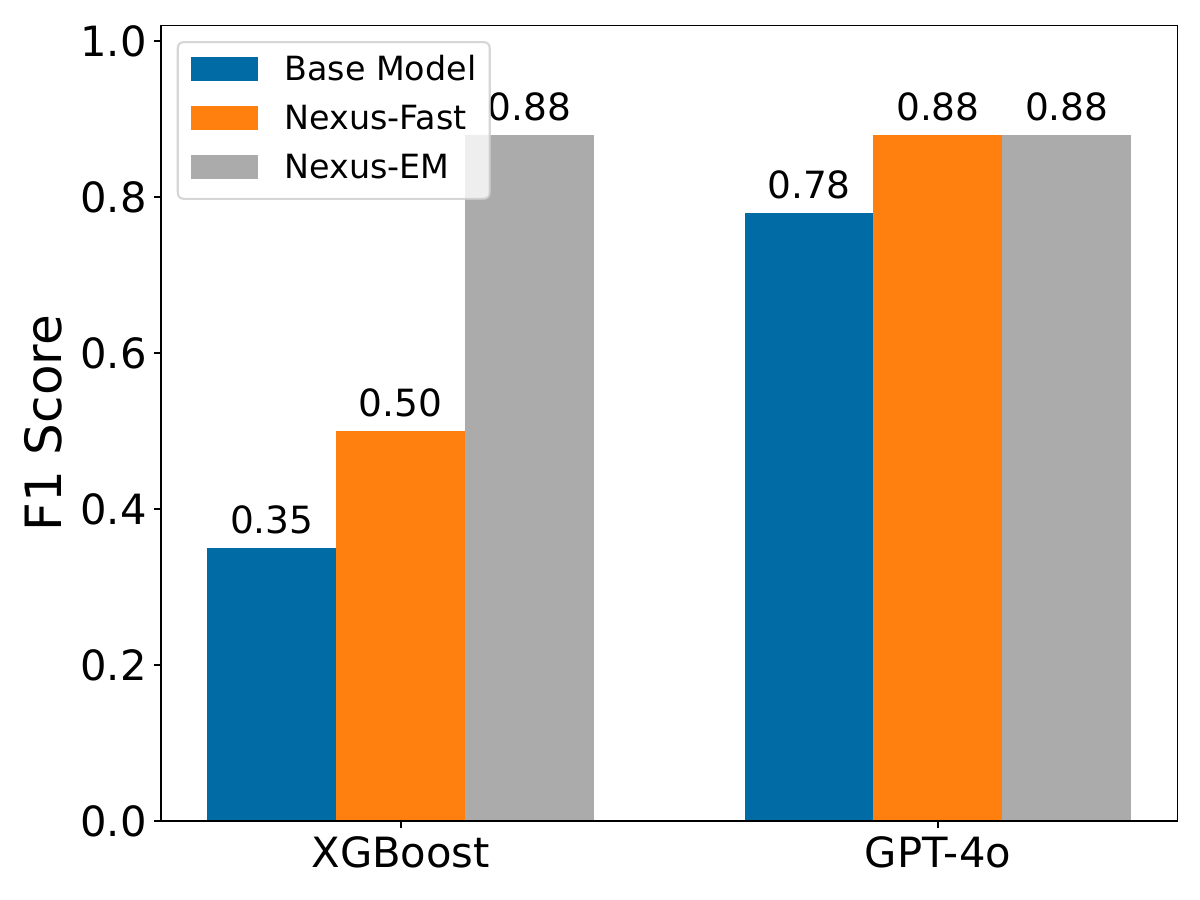}
      \caption*{TPC-H}
    \end{minipage} \hfill
    \begin{minipage}[t]{0.245\textwidth}
      \centering
      \includegraphics[width=\columnwidth]{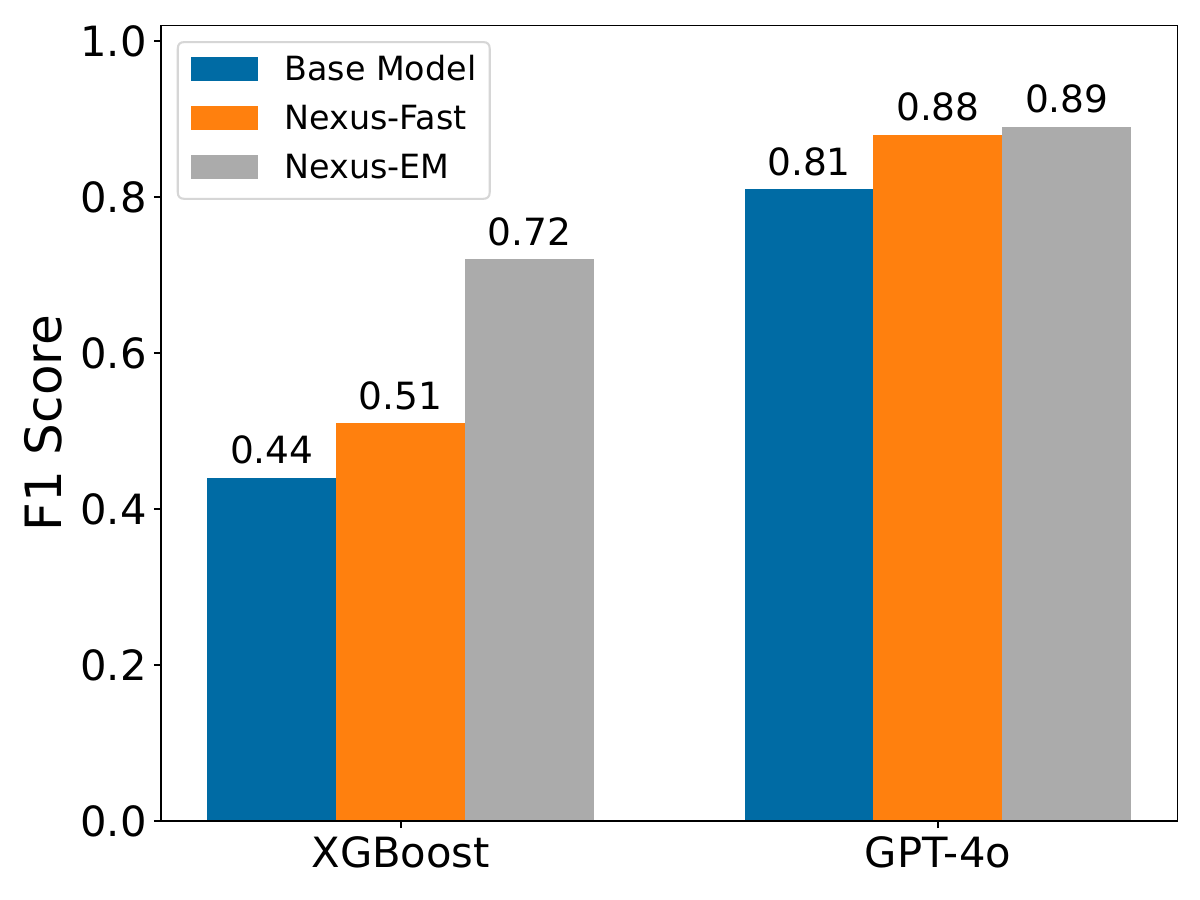}
      \caption*{TPC-DS}
    \end{minipage} \hfill
    \begin{minipage}[t]{0.245\textwidth}
      \centering
      \includegraphics[width=\columnwidth]{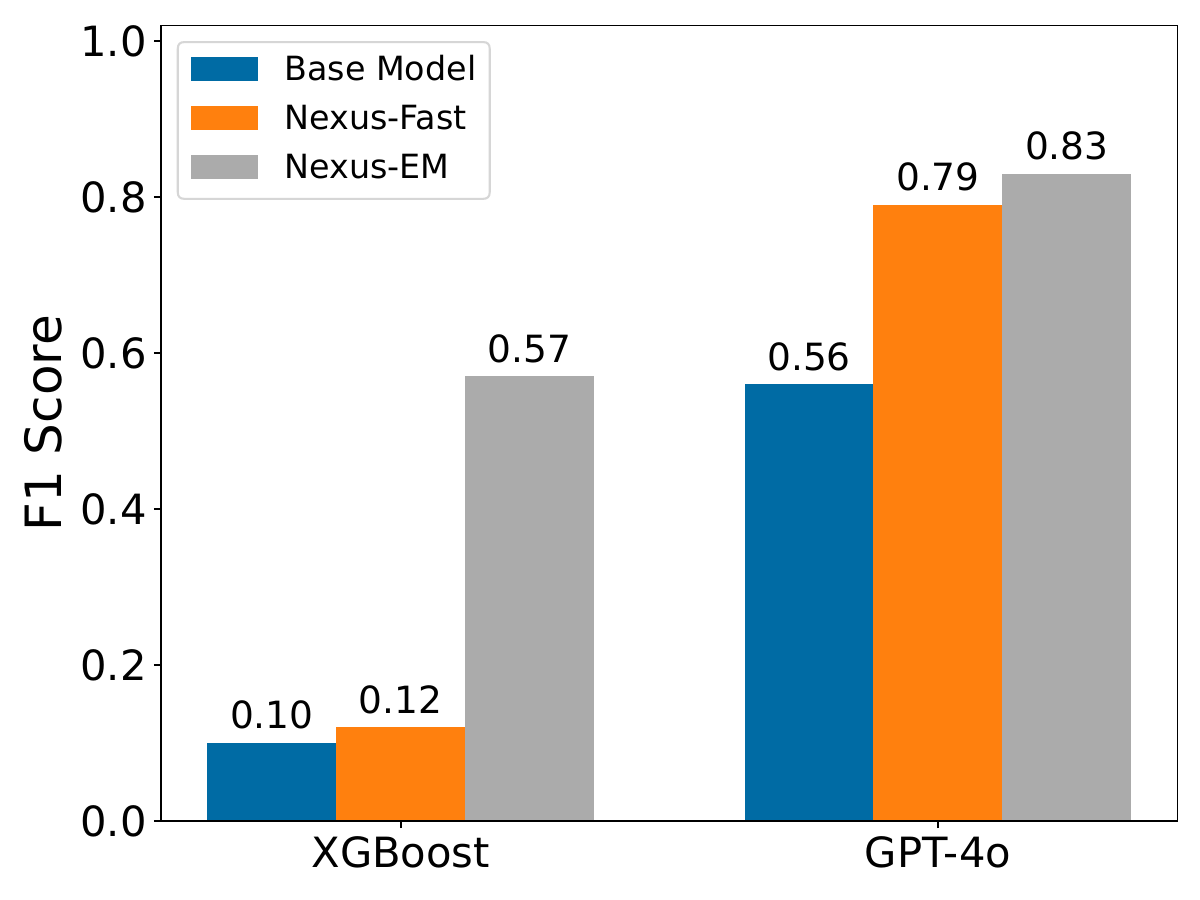}
      \caption*{BIRD-SQL}
    \end{minipage} \hfill
    \begin{minipage}[t]{0.245\textwidth}
      \centering
      \includegraphics[width=\columnwidth]{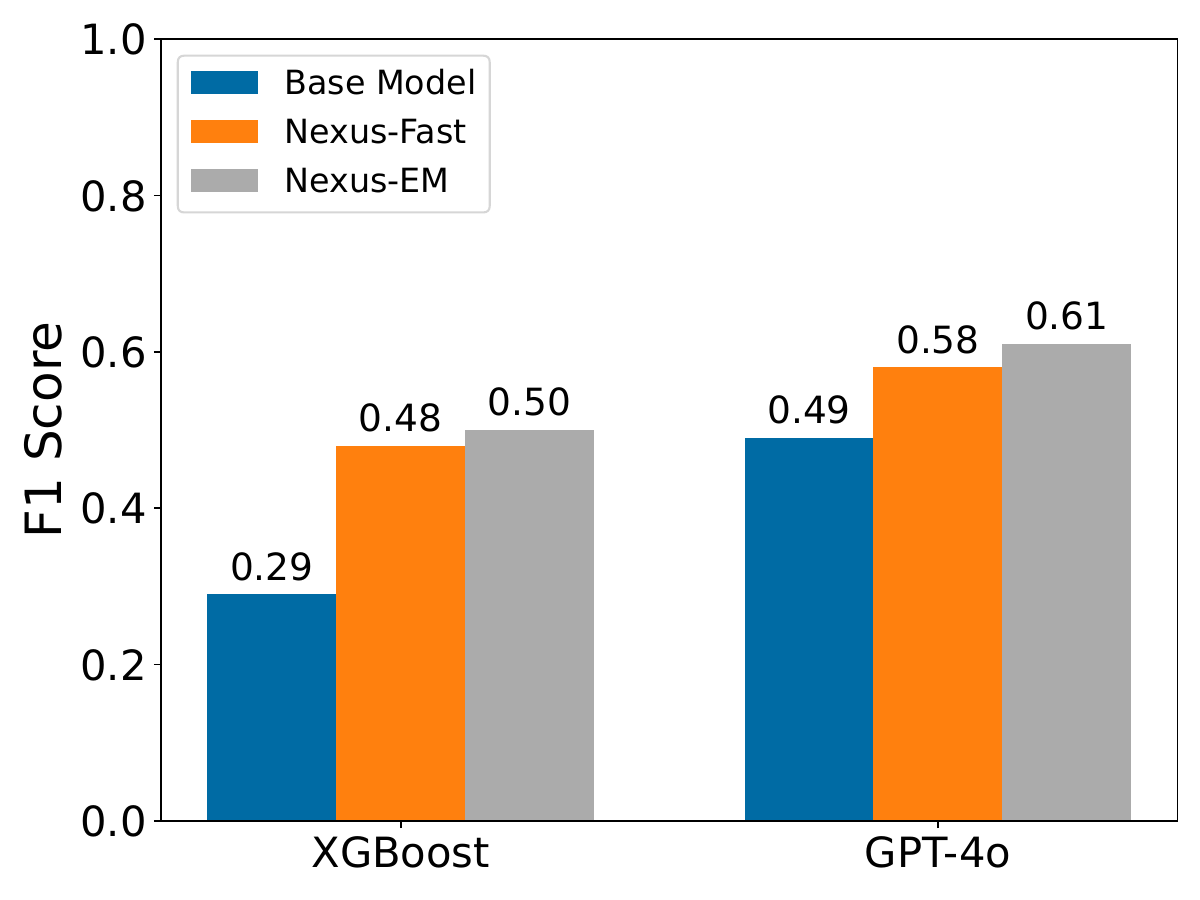}
      \caption*{REAL}
    \end{minipage}
    \caption{F1 score comparison of base models, \sysname-Fast, and \sysname-EM across four datasets.}
    \label{fig:fast_mode_quality}
  \end{figure*}

\subsection{Quality Comparison}
  Table~\ref{tab:quality_comparison} presents the F1 score, precision and recall for all metadata-only approaches across the four datasets. \sysname consistently outperforms all baselines by a significant margin. For instance, \sysname-GPT-4o (\sysname with GPT-4o priors) surpasses Auto-BI, the previous SOTA, by 32, 37, 51, and 53 F1 points on the TPC-H, TPC-DS, BIRD-SQL and REAL datasets, respectively. Similarly, SemPy lags behind \sysname-GPT-4o by 47, 46, 54, and 35 F1 points on these datasets.

  XGBoost, pretrained on the SchemaPile-Perm dataset with only metadata features, struggles to generalize to the test datasets, yielding the lowest F1 scores on TPC-H and BIRD-SQL. In contrast, even with these weak priors, \sysname-XGBoost significantly improves the F1 score by 53 points on TPC-H, 28 on TPC-DS, 47 on BIRD-SQL, and 21 on REAL. This substantial improvement is attributed to \sysname's iterative optimization (the EM algorithm in Algorithm~\ref{alg:em_optimization}), which leverages the low-rank structure of join graphs and LLM-based column entity type checks. 

  \sysname also improves upon GPT-4o priors, even when the model shows strong out-of-the-box results. GPT-4o's strong result quality on the public TPC-H and TPC-DS benchmarks is expected, as it likely memorized their join ground truth during training. While BIRD-SQL is also public, combining all 11 databases into a single dataset causes GPT-4o to generate a non-trivial number of false positives (evidenced by its lower precision relative to its nearly perfect recall). For the unseen REAL dataset, GPT-4o correctly predicts some true joins but suffers from a high false positive rate.

  The pre-trained model serves as a modular component within \sysname. While we use GPT-4o to exemplify a strong prior, it can be replaced by any advanced ML model. The results of \sysname-XGBoost and \sysname-GPT-4o highlight the robustness of our framework: while a more accurate prior naturally yields better results, \sysname delivers adequate quality even when initialized with a significantly weaker model. In particular, \sysname-XGBoost matches the high result quality of \sysname-GPT-4o on TPC-H, even though its underlying XGBoost prior is substantially weaker than the GPT-4o prior (F1 scores of 0.35 vs. 0.78).

  Auto-BI and SemPy underperform on all datasets. This outcome is not surprising, as both methods were originally designed assuming access to data values. The lack of data access prevents the use of inclusion dependency checks for pruning candidates, leading to higher false positive rates. Additionally, Auto-BI's performance relies on a reasonably effective base model, and its results are significantly compromised by the absence of features derived from data values. As shown in Section~\ref{subsec:when_data_available}, both methods demonstrate much-improved results when granted full data access. However, for the specific task of inferring a join graph from metadata alone, neither Auto-BI nor SemPy is competitive.

  Finally, we observe a consistent trend: as dataset complexity increases, from simple single-schema settings (TPC-H and TPC-DS) to multi-source scenarios (BIRD-SQL) and complex production environments (REAL), result quality declines across all approaches. Specifically, all methods suffer from high false positive rates (and consequently low precision) in multi-source datasets. Rather than painting a rosy picture based on synthetic data, the REAL dataset provides a true measure of what existing methods can achieve in actual production settings (datasets from multiple sources residing in data lakes or lakehouses).

  \begin{figure*}[ht!]
    \begin{minipage}{0.265\textwidth}
      \centering
      \includegraphics[width=\columnwidth]{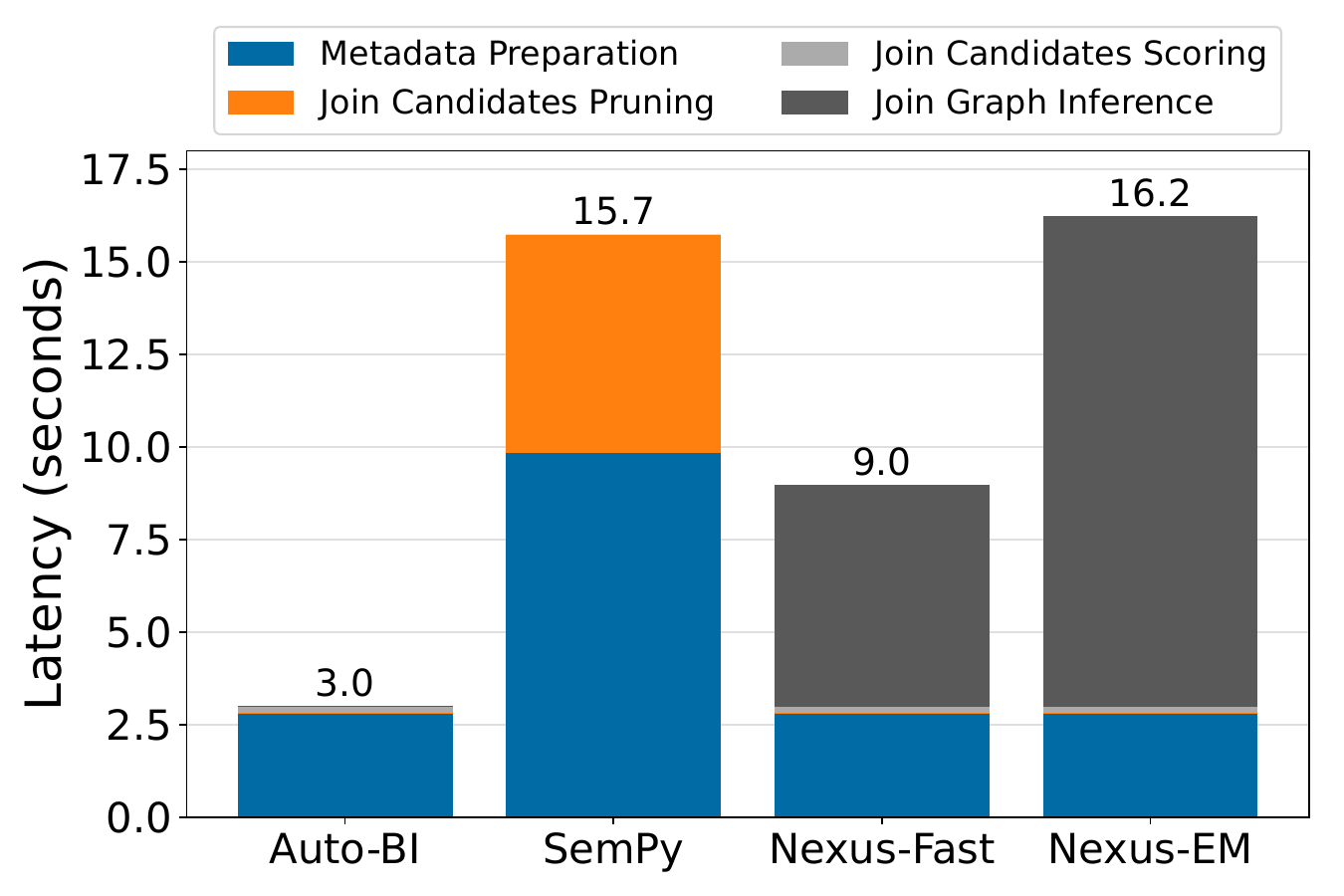}
      \caption*{TPC-H}
    \end{minipage} \hfill
    \begin{minipage}{0.24\textwidth}
      \centering
      \includegraphics[width=\columnwidth]{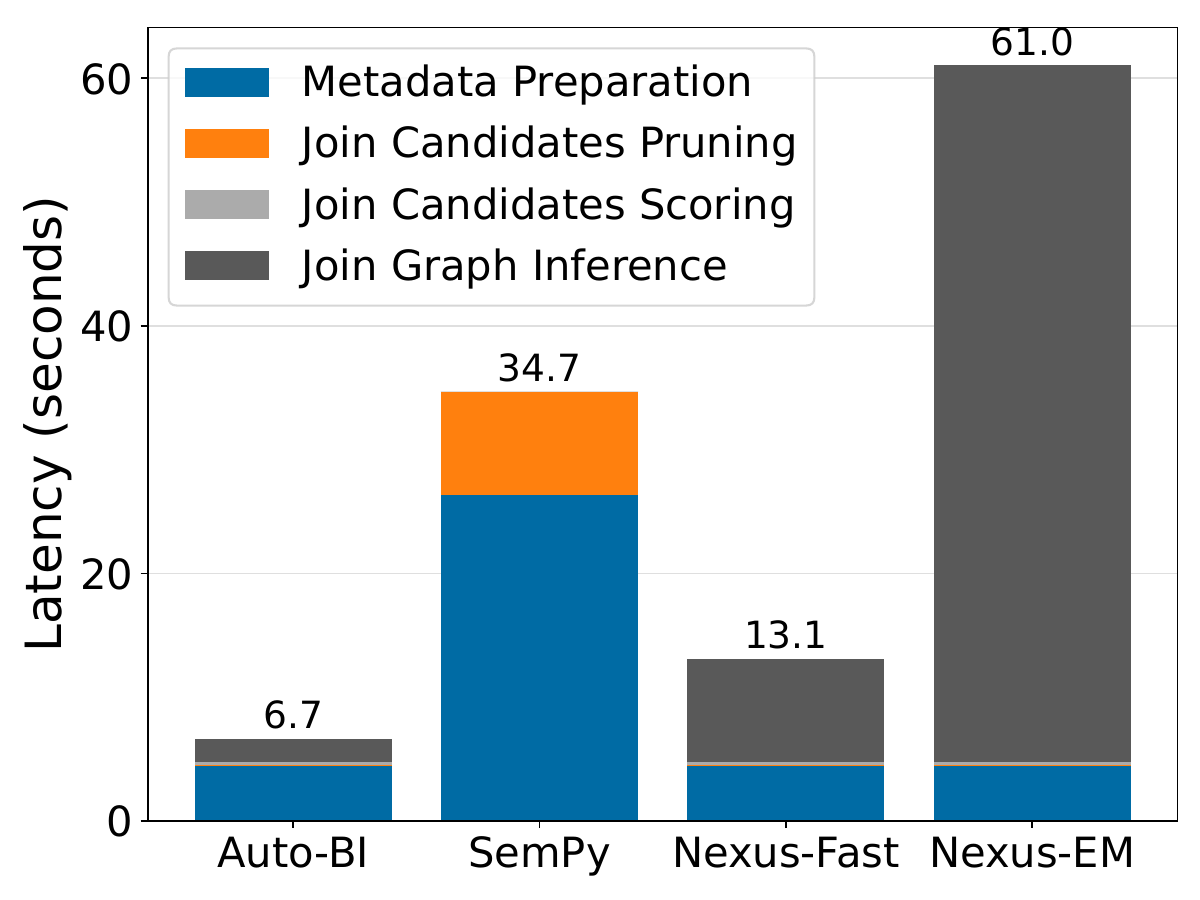}
      \caption*{TPC-DS}
    \end{minipage} \hfill
     \begin{minipage}{0.24\textwidth}
      \centering
      \includegraphics[width=\columnwidth]{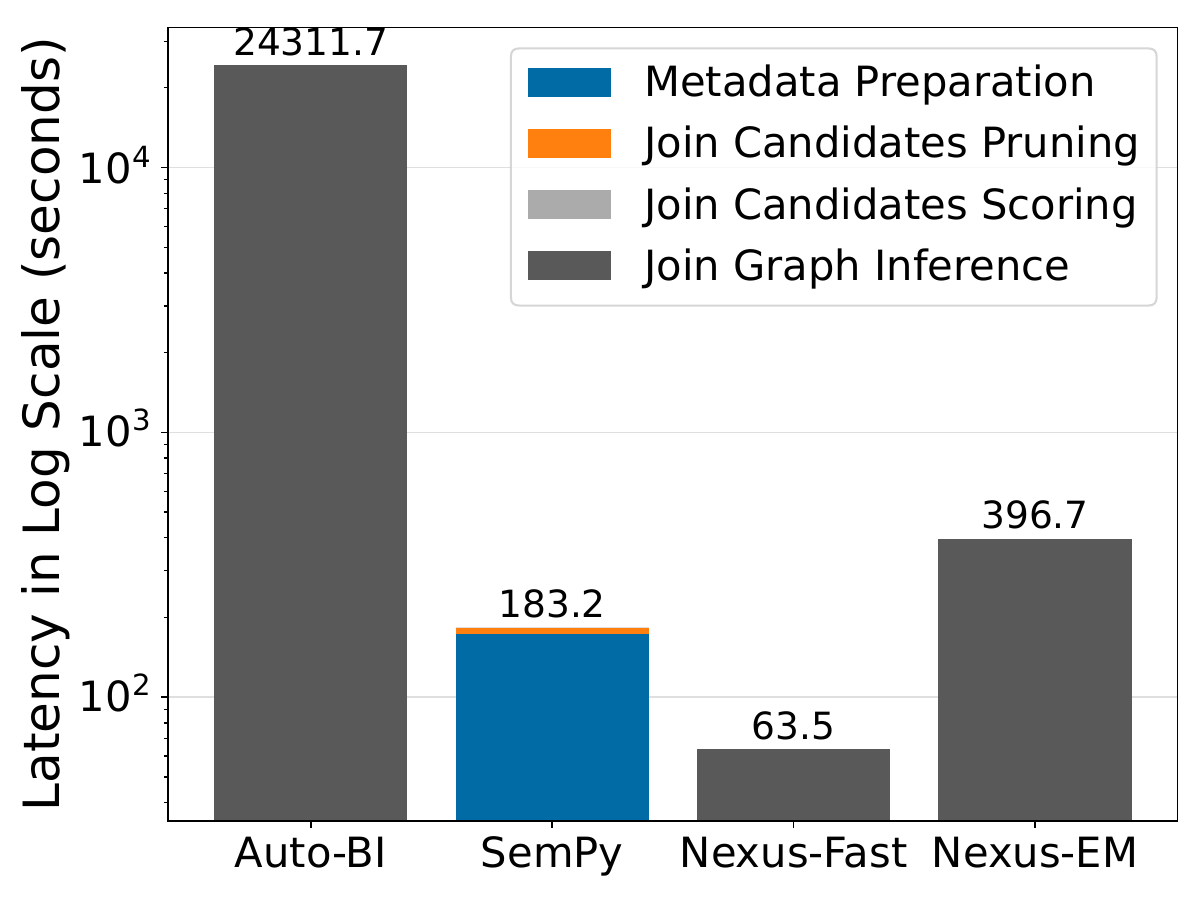}
      \caption*{BIRD-SQL}
    \end{minipage}
    \begin{minipage}{0.24\textwidth}
      \centering
      \includegraphics[width=\columnwidth]{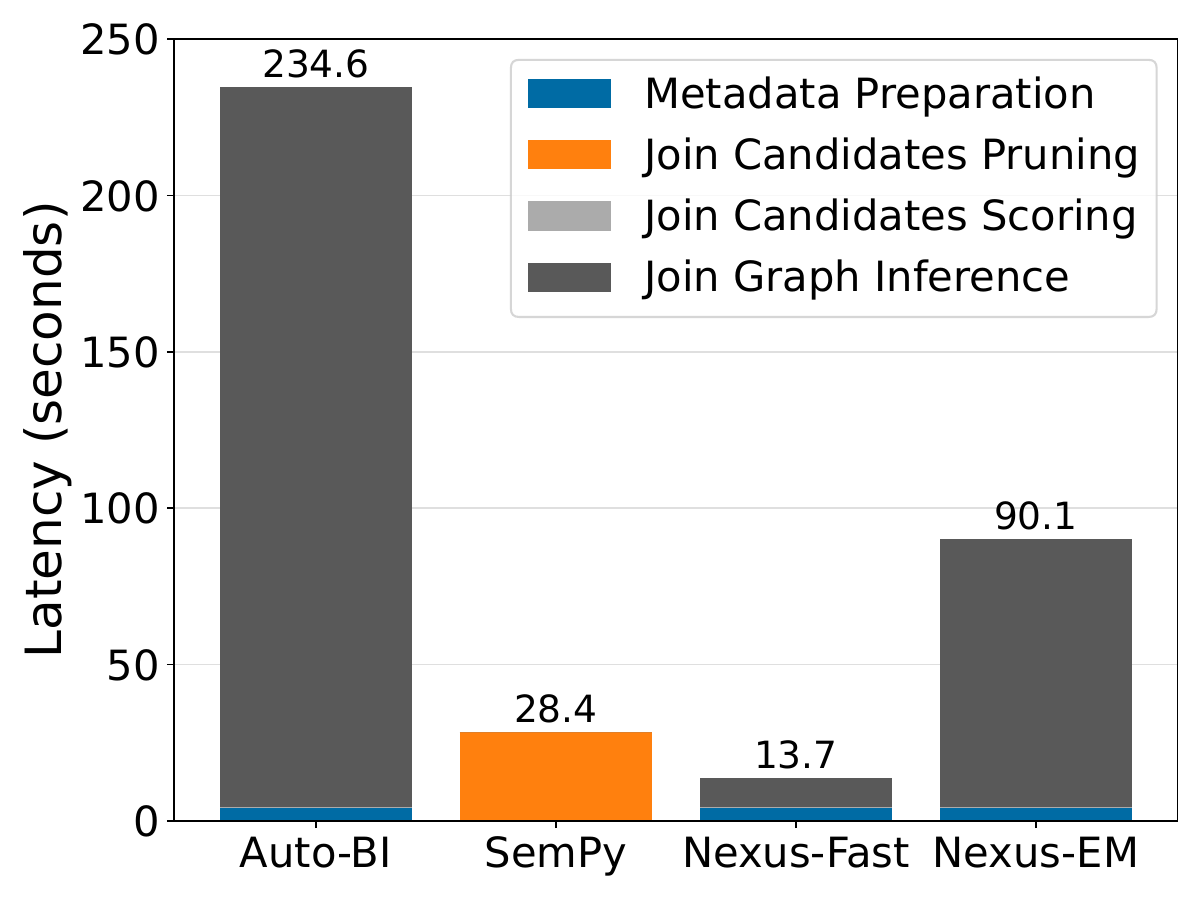}
      \caption*{REAL}
    \end{minipage}
    \caption{Efficiency comparison on TPC-H, TPC-DS, BIRD-SQL, and REAL datasets. Auto-BI shares Nexus's implementations of metadata preparation, join candidates pruning and join candidates scoring.}
    \label{fig:efficiency_comparison}
    \vspace{-3mm}
  \end{figure*}

\subsection{Fast Mode Inference}\label{subsec:exp_fast_mode_inference}
  By default, \sysname leverages LLMs within its EM algorithm to infer semantic entity types of columns and perform soft type matching. While powerful, these LLM API calls can incur increased latency and considerable monetary costs. For latency- or budget-sensitive scenarios, we offer \sysname-Fast, a fast mode of \sysname that exclusively employs low-rank matrix completion (no iterative EM algorithm) on the core submatrix (as discussed in Section~\ref{subsec:fast_mode}).
  
  For clarity in our comparison, we refer to our default algorithm with EM iterations as \sysname-EM to distinguish from \sysname-Fast. In Figure~\ref{fig:fast_mode_quality}, we compare the F1 score of \sysname-Fast and \sysname-EM, using XGBoost and GPT-4o as base models for scoring join candidates. Overall, \sysname-Fast can significantly improve result quality of both base models across four datasets while \sysname-EM achieves the best results.
  
  When a weak base model is used, like XGBoost, \sysname-EM empowered by LLM delivers stronger results than \sysname-Fast, although it has a longer runtime (as shown in the next section). Specifically, \sysname-EM outperforms \sysname-Fast by 38, 21, and 45 F1 points on the TPC-H, TPC-DS, and BIRD-SQL datasets, respectively. When the weak model provides poor priors, as seen with BIRD-SQL, \sysname-Fast barely improves result quality by 2 F1 points. Nevertheless, when the same base model has some predictive value as with TPC-H, TPC-DS, and REAL, \sysname-Fast achieves a boost of 15, 7 and 19 F1 points, respectively.
  
  In contrast, when a strong base model like GPT-4o is employed, \sysname-Fast yields result quality comparable to \sysname-EM. The quality gap significantly shrinks, narrowing to only 1 F1 point on TPC-DS, 4 on BIRD-SQL, 3 on REAL, and no difference on TPC-H. Compared to the base model alone, \sysname-Fast yields improvements of 10, 7, 23, and 9 F1 points on these four datasets. The next section will elaborate on the latency advantages of \sysname-Fast.

\subsection{Efficiency Comparison}
  Figure~\ref{fig:efficiency_comparison} shows the efficiency comparison of the four methods across all datasets. We report the runtime of \sysname-Fast and \sysname-EM using the XGBoost base model here. 

  We decompose end-to-end latency into four key components: metadata preparation, join candidates pruning, join candidates scoring, and join graph inference, as illustrated in our pipeline (Figure~\ref{fig:solution_overview}). Auto-BI, \sysname-Fast, and \sysname-EM incorporate all four components, whereas SemPy only involves metadata preparation and join candidates pruning. To ensure a fair comparison in our metadata-only adaptation, we replaced the first three components of Auto-BI with the more efficient implementations from \sysname.
  
  Among all methods, \sysname-Fast demonstrates competitive efficiency, particularly on multi-source datasets with large numbers of tables and columns. On the BIRD-SQL and REAL datasets, \sysname-Fast is more than 6x faster than \sysname-EM. This performance gap is expected, as \sysname-Fast exclusively performs low-rank matrix completion on the core submatrix, whereas \sysname-EM incurs the additional cost of LLM API calls for semantic column type checks in each iteration. The latency for Nexus-EM also depends on the number of iterations we run the algorithm for. Here, we run \sysname-EM for five iterations, which is sufficient to achieve the best result on each dataset. Nevertheless, \sysname-EM completes join detection within a few minutes even on large multi-source datasets. Auto-BI exhibits the highest latency on BIRD-SQL and REAL datasets, running 382x and 17x slower than \sysname-Fast, respectively. This suggests that Auto-BI's graph-based optimization technique struggles to scale to larger multi-source datasets.

  Together, Figure~\ref{fig:fast_mode_quality} and Figure~\ref{fig:efficiency_comparison} demonstrate that \sysname-Fast offers an attractive trade-off between quality and efficiency when utilizing a reasonably accurate base model.

  \begin{figure*}[ht]
    \begin{minipage}{0.245\textwidth}
      \centering
      \includegraphics[width=\columnwidth]{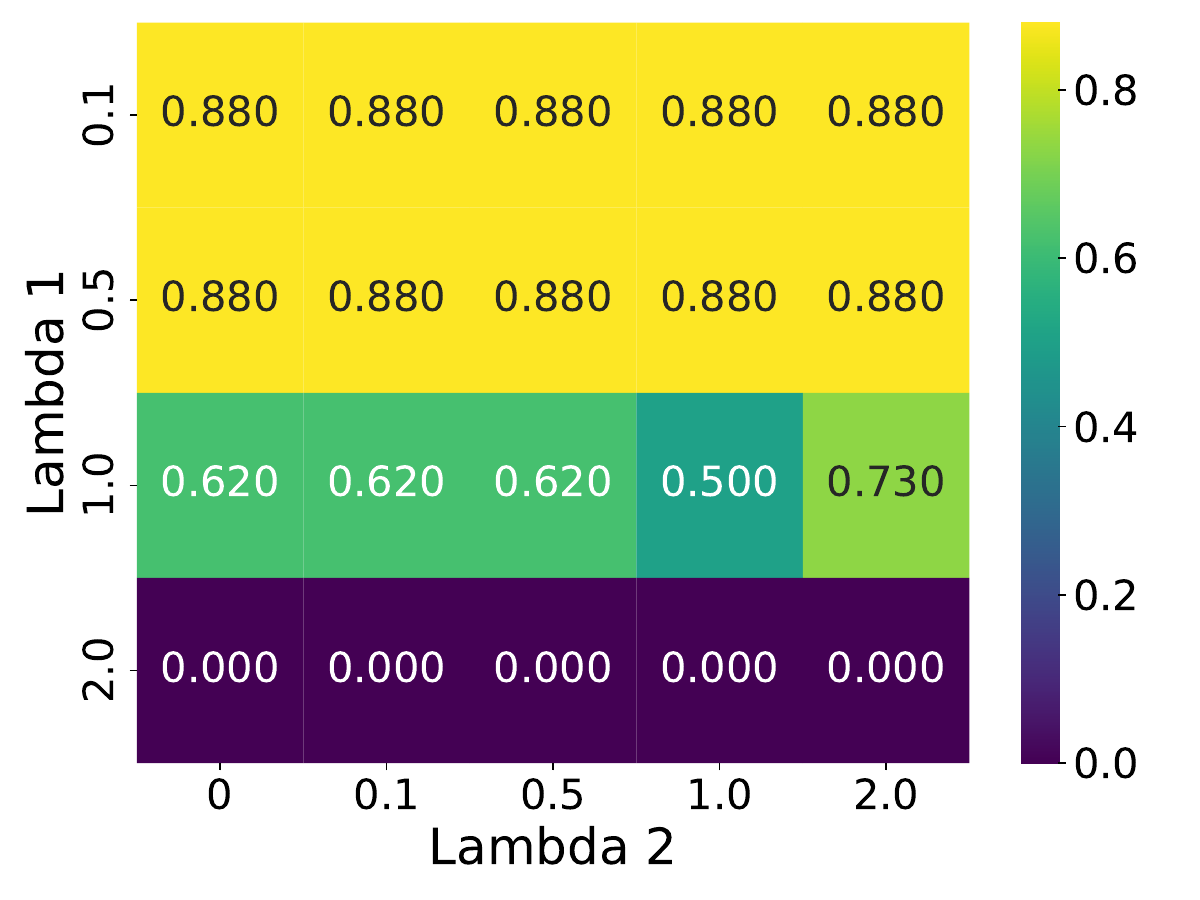}
      \caption*{TPC-H}
    \end{minipage} \hfill
    \begin{minipage}{0.245\textwidth}
      \centering
      \includegraphics[width=\columnwidth]{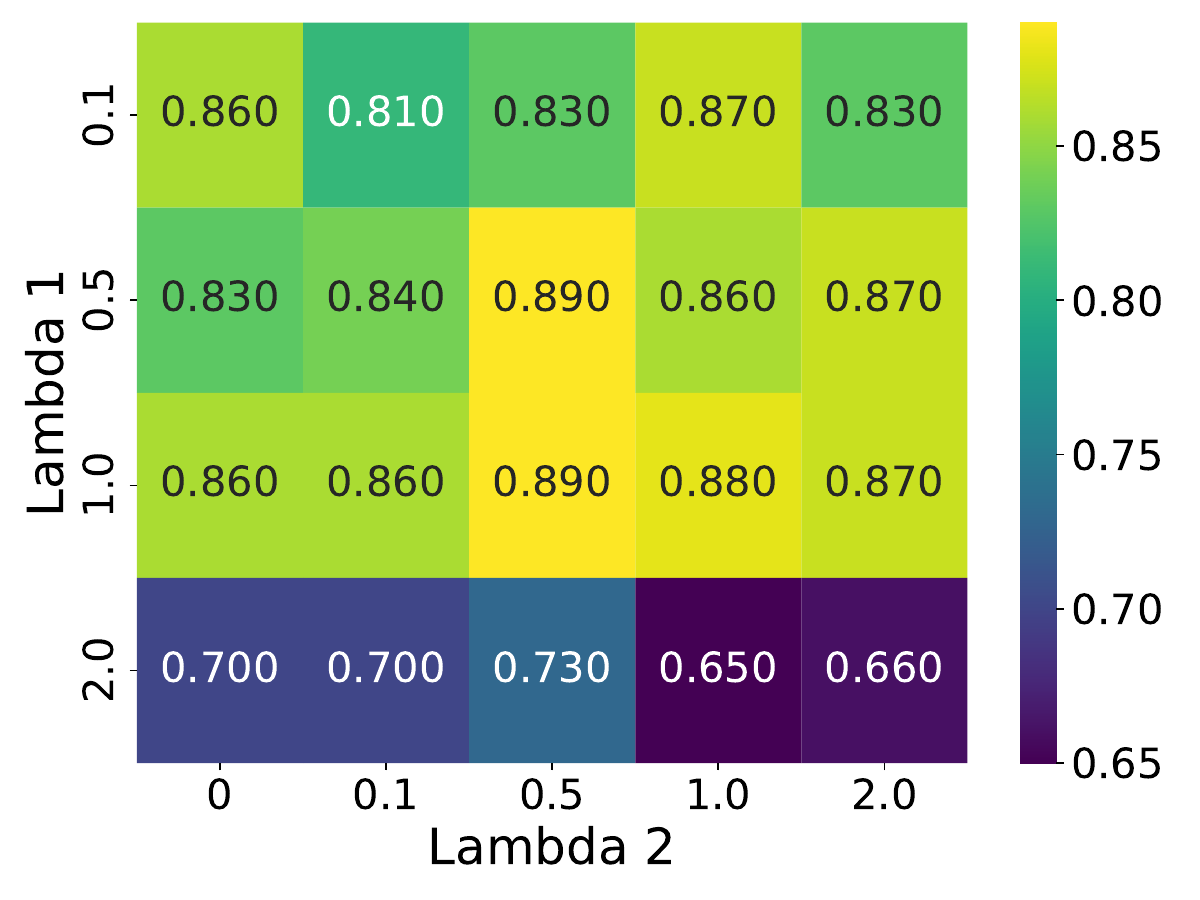}
      \caption*{TPC-DS}
    \end{minipage}
    \begin{minipage}{0.245\textwidth}
      \centering
      \includegraphics[width=\columnwidth]{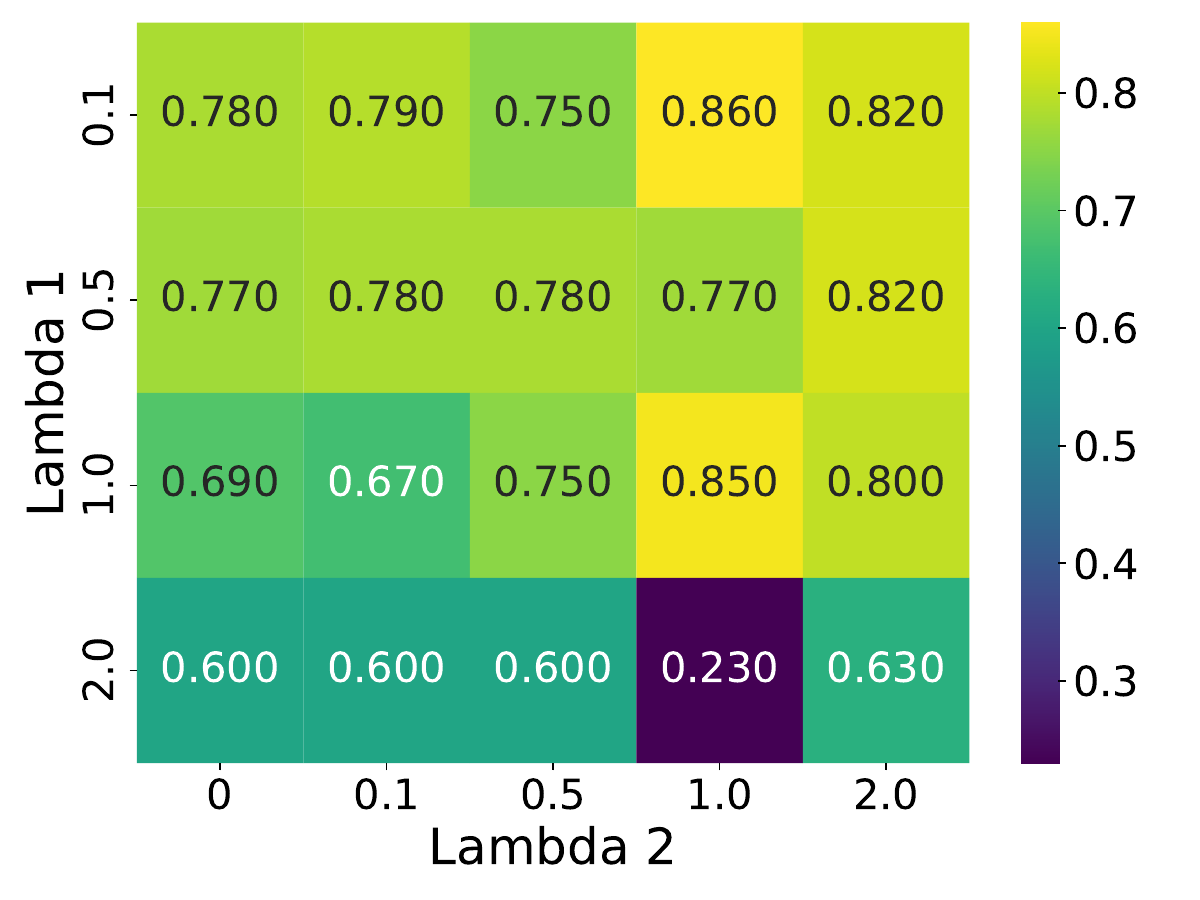}
      \caption*{BIRD-SQL}
    \end{minipage} \hfill
    \begin{minipage}{0.245\textwidth}
      \centering
      \includegraphics[width=\columnwidth]{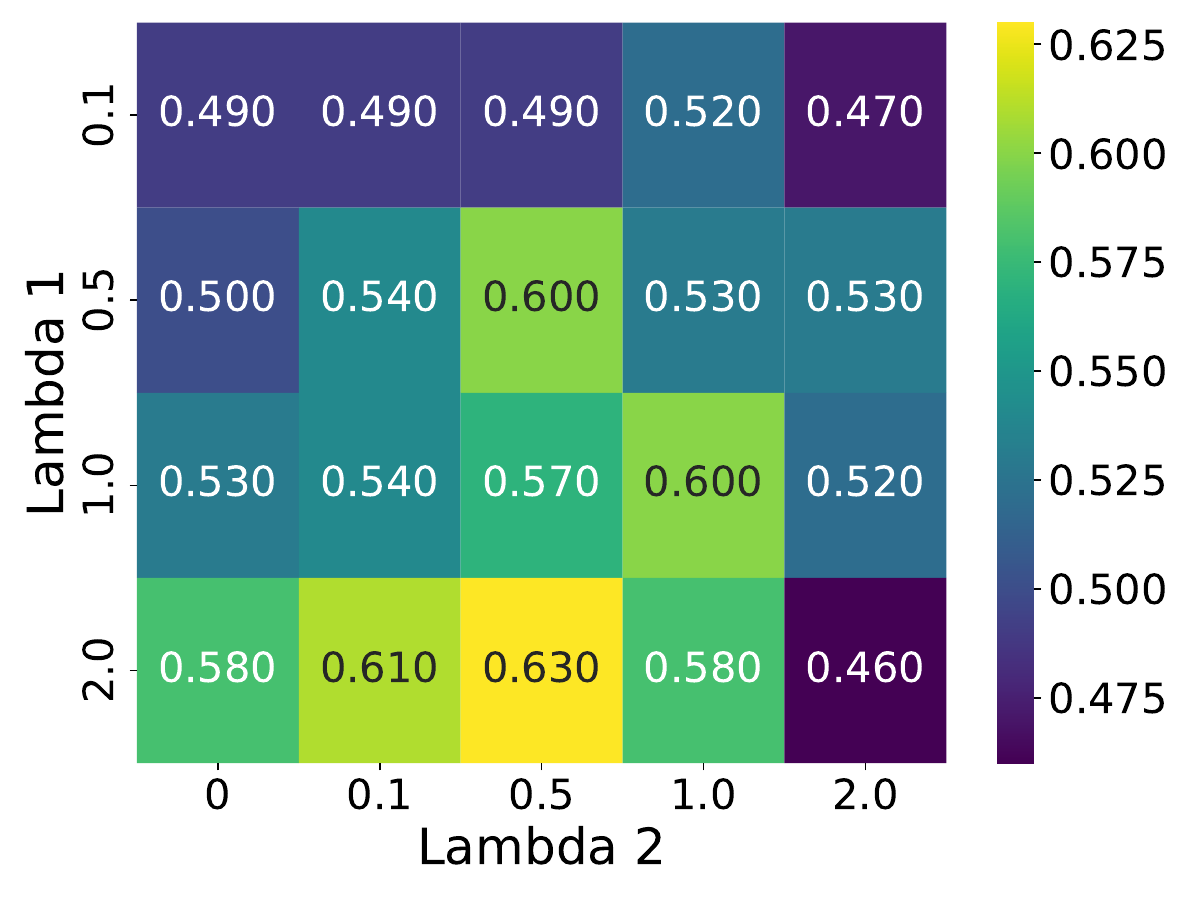}
      \caption*{REAL}
    \end{minipage}
    \caption{F1 score heat maps of \sysname-GPT-4o w.r.t loss weights $\lambda_{1}$ and $\lambda_{2}$ in the low-rank matrix completion objective function.}
    \label{fig:ablation_lambda}
  \end{figure*}

  \begin{figure*}[ht!]
    \begin{minipage}{0.245\textwidth}
      \centering
      \includegraphics[width=\columnwidth]{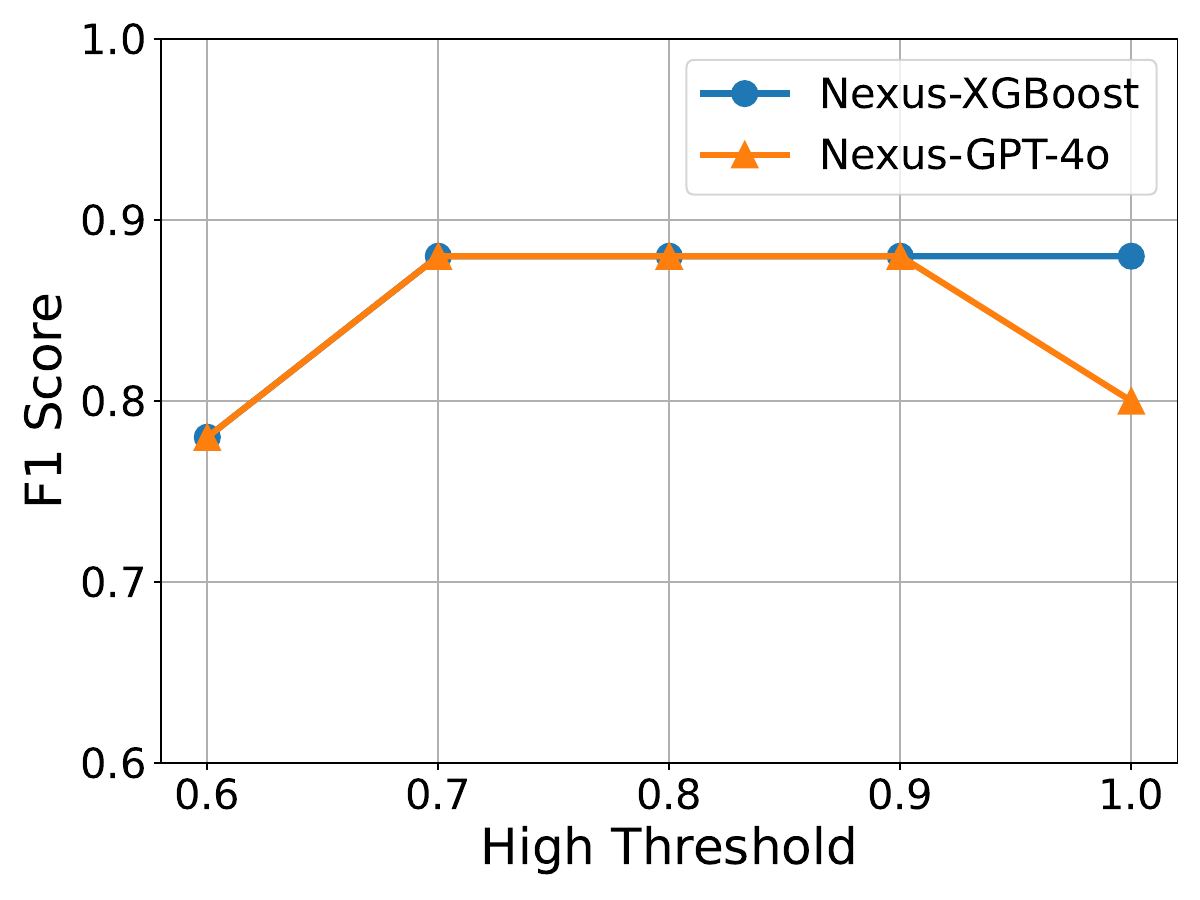}
      \caption*{TPC-H}
    \end{minipage} \hfill
    \begin{minipage}{0.245\textwidth}
      \centering
      \includegraphics[width=\columnwidth]{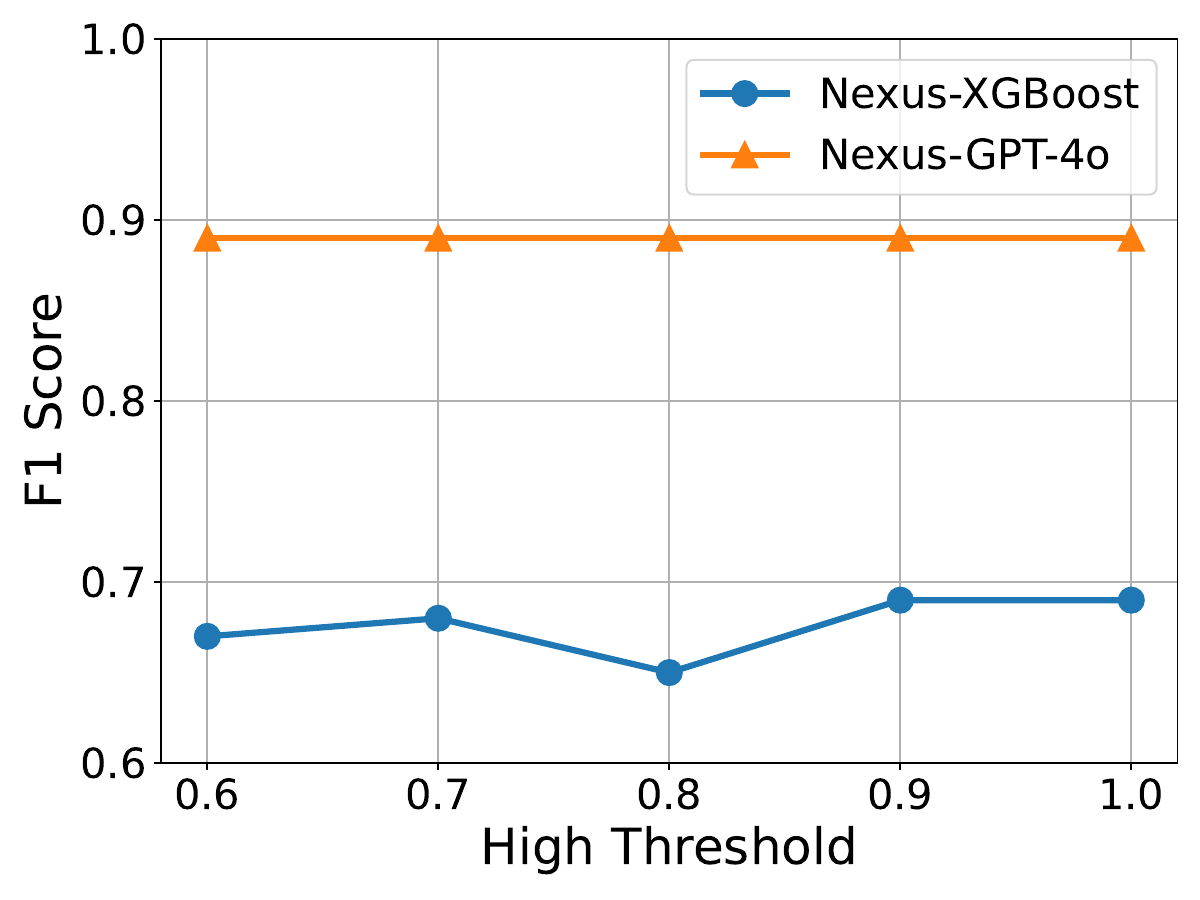}
      \caption*{TPC-DS}
    \end{minipage} \hfill
     \begin{minipage}{0.245\textwidth}
      \centering
      \includegraphics[width=\columnwidth]{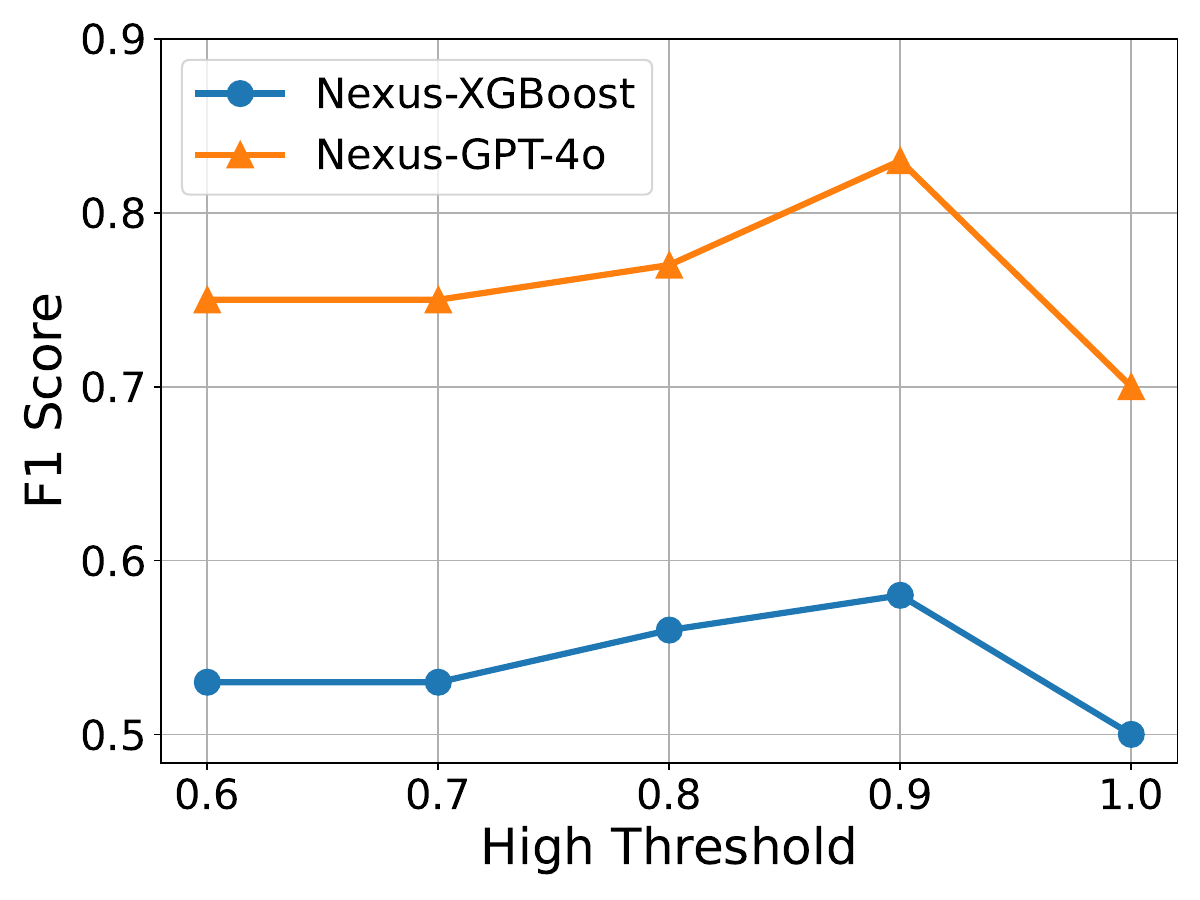}
      \caption*{BIRD-SQL}
    \end{minipage} \hfill
    \begin{minipage}{0.245\textwidth}
      \centering
      \includegraphics[width=\columnwidth]{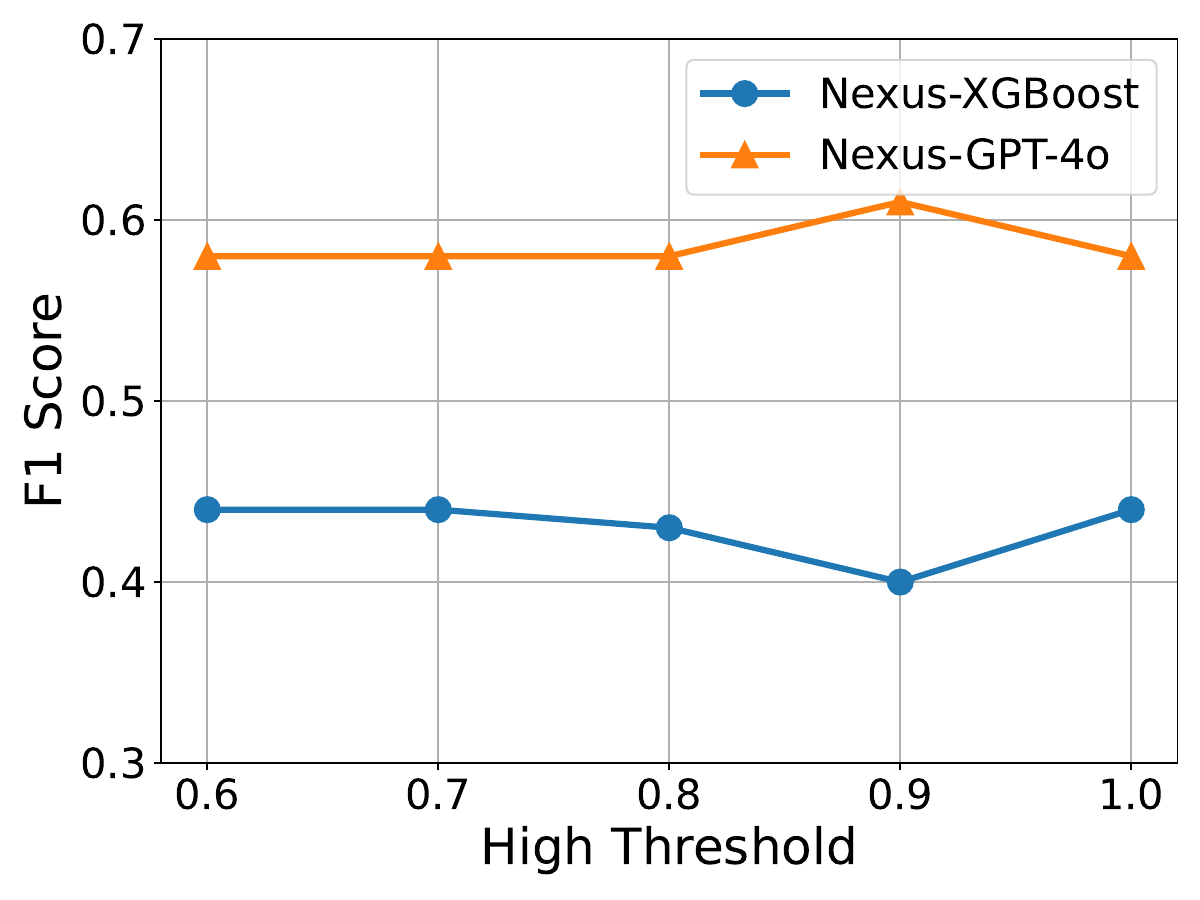}
      \caption*{REAL}
    \end{minipage}
    \caption{F1 score w.r.t high\_threshold in the \sysname-EM algorithm.}
    \label{fig:ablation_high_threshold}
  \end{figure*}

\subsection{Ablation Study of Hyperparameters}
  We conduct an ablation study to analyze the impact of key hyperparameters on the F1 score.

  \subsubsection{Weights $\lambda_{1}$ and $\lambda_{2}$ in the LRMC objective function} The parameters $\lambda_{1}$ and $\lambda_{2}$ balance between low-rank regularization and sparsity regularization in our optimization objective of low-rank matrix completion (equation~\ref{eq:full_objective}). We evaluate a grid of 20 combinations, comprising four values for $\lambda_{1}$ (0.1, 0.5, 1.0, 2.0) and five values for $\lambda_{2}$ (0, 0.1, 0.5, 1.0, 2.0). Figure~\ref{fig:ablation_lambda} presents the F1 score heat maps for \sysname-GPT-4o. Overall, \sysname-GPT-4o yields optimal results with moderate regularization strength. Specifically, $\lambda_{1} \in \{0.5,1.0\}$ performs best on TPC-H, TPC-DS, and BIRD-SQL, whereas a larger $\lambda_{1}$ is optimal for REAL. This aligns with the dataset properties: REAL has the lowest normalized rank (0.02 vs. 0.20 for TPC-H, as shown in Table~\ref{tab:dataset_stats}), necessitating stronger low-rank regularization. Due to space constraints, we omit heatmaps for \sysname-XGBoost where it exhibits a similar pattern.

  \subsubsection{Thresholds in the EM algorithm} Our EM algorithm introduces two parameters: $low\_threshold$ and $high\_threshold$. Since we observe that base models exhibit a low false negative rate, we fix $low\_threshold$ to 0.5, which is also the decision threshold of base models. In other words, if a base model predicts a candidate pair negative, we bypass the expensive LLM semantic check for this pair in the EM algorithm. This allows \sysname-EM to focus on pairs predicted as positive, ensuring execution completes within minutes, even for larger datasets. On the other hand, $high\_threshold$ controls how soon we mark a type-compatible candidate pair as a definitive positive. Figure~\ref{fig:ablation_high_threshold} illustrates the F1 score sensitivity to $high\_threshold$ for both \sysname-XGBoost and \sysname-GPT-4o. Overall, both methods are relatively robust to the change of $high\_threshold$.

\begin{figure*}[ht]
    \begin{minipage}[t]{0.24\textwidth}
      \centering
      \includegraphics[width=\columnwidth]{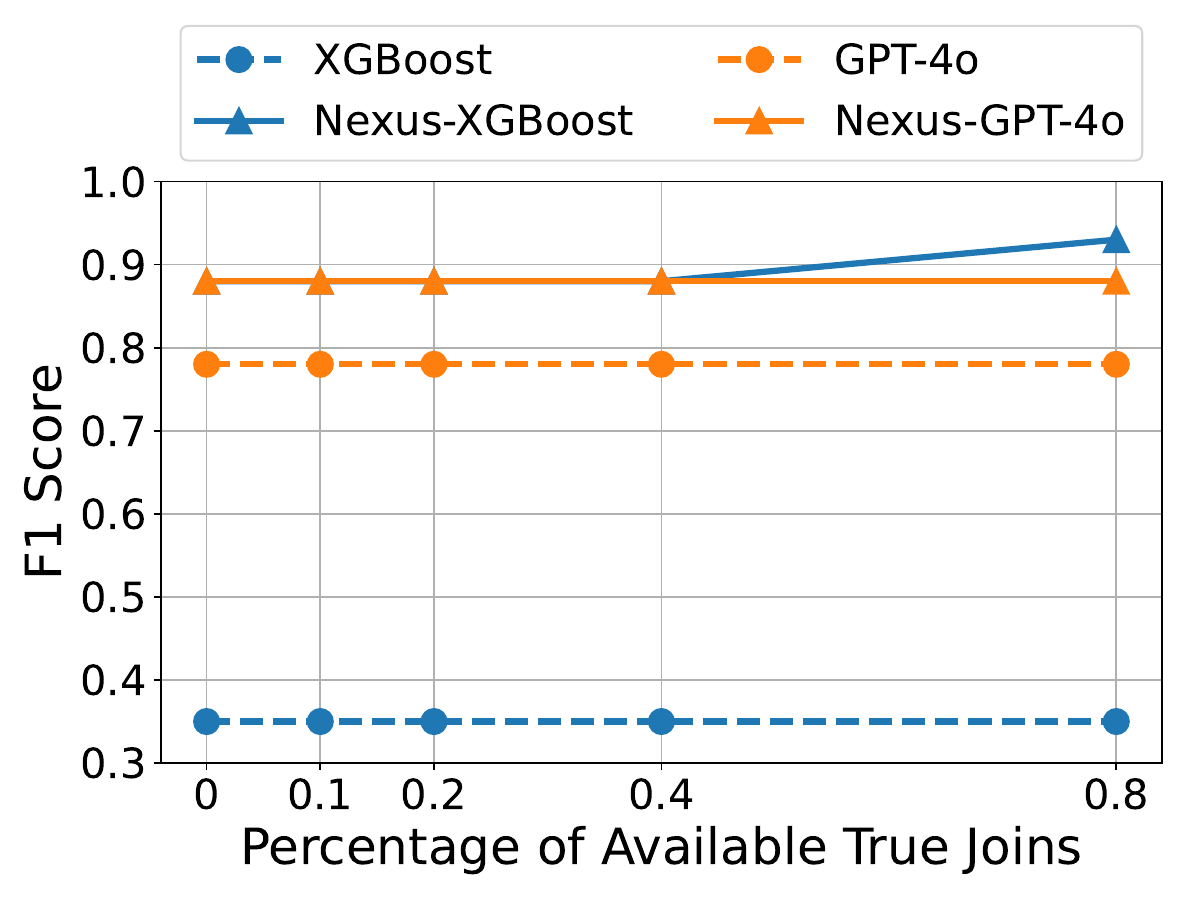}
      \caption*{TPC-H}
    \end{minipage} \hfill
    \begin{minipage}[t]{0.24\textwidth}
      \centering
      \includegraphics[width=\columnwidth]{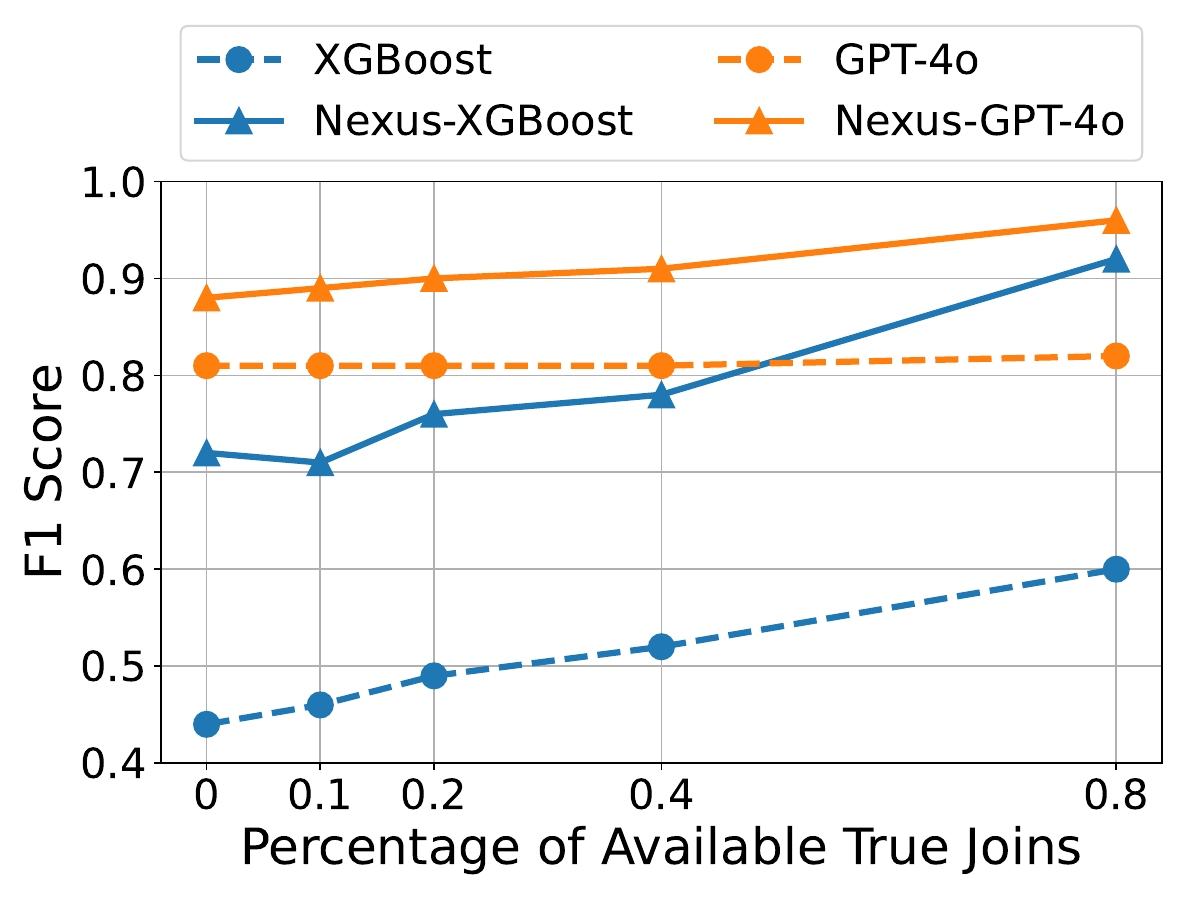}
      \caption*{TPC-DS}
    \end{minipage} \hfill
    \begin{minipage}[t]{0.24\textwidth}
      \centering
      \includegraphics[width=\columnwidth]{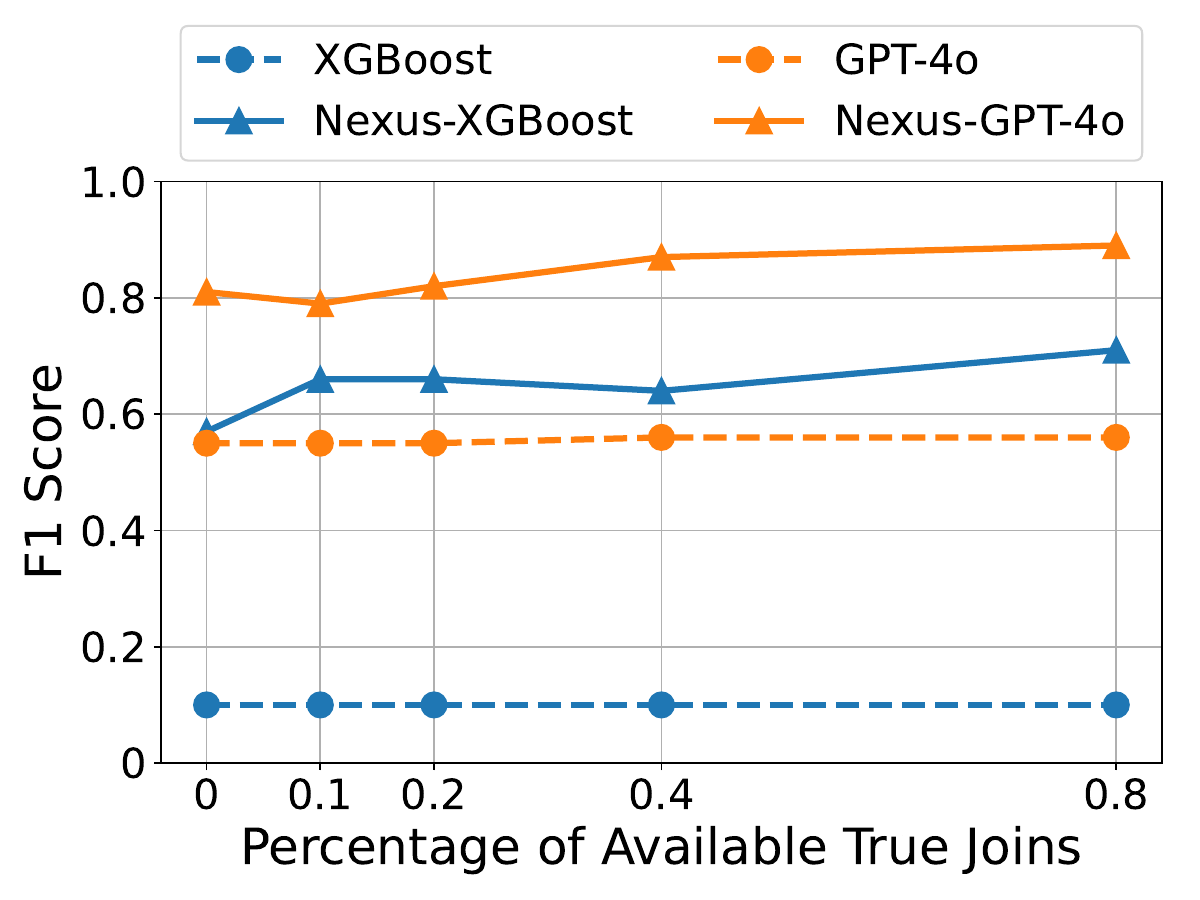}
      \caption*{BIRD-SQL}
    \end{minipage} \hfill
    \begin{minipage}[t]{0.24\textwidth}
      \centering
      \includegraphics[width=\columnwidth]{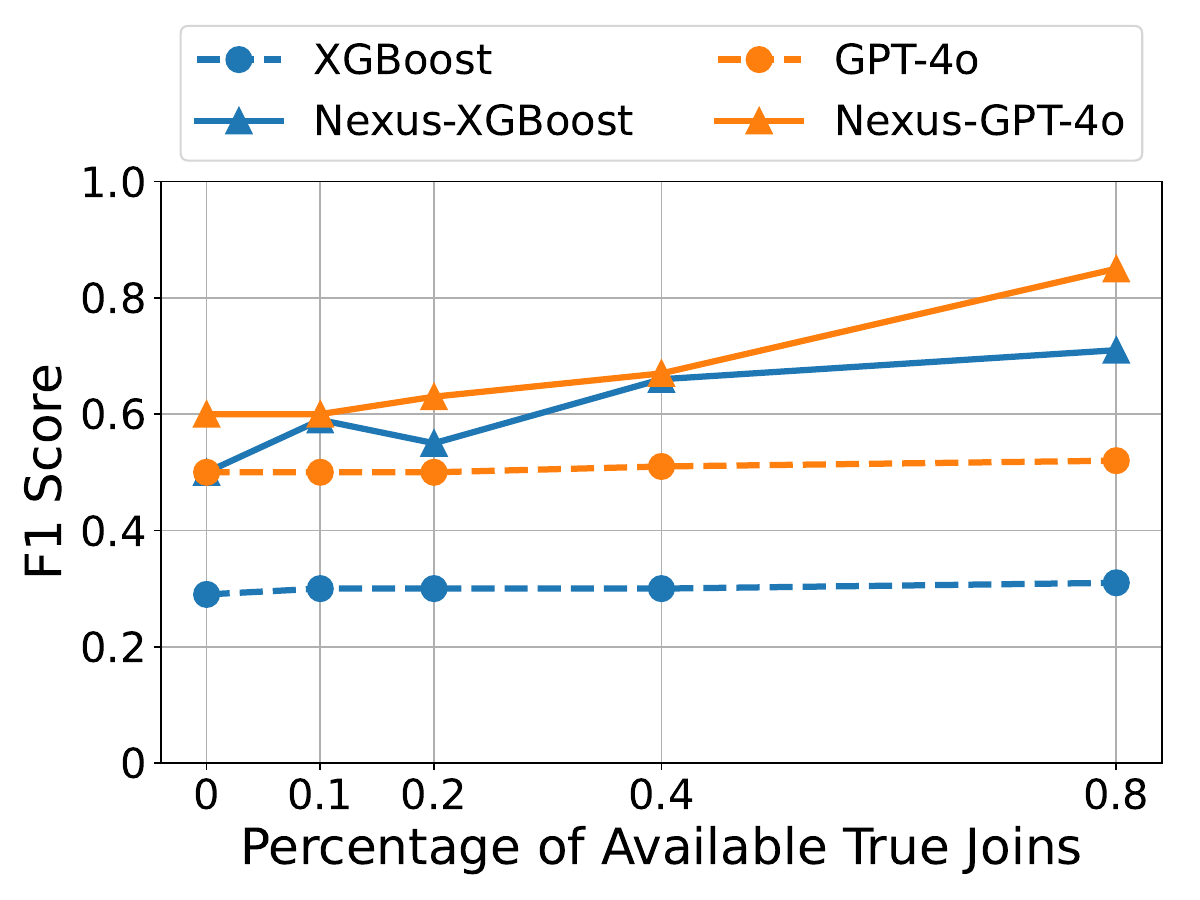}
      \caption*{REAL}
    \end{minipage}
    \caption{F1 score of base models (dashed lines) and \sysname-EM (solid lines) w.r.t the percentage of available true joins.}
    \label{fig:quality_vs_sample_ratio}
  \end{figure*}

\subsection{When Query Logs are Available}\label{subsec:when_query_log_available}
  In practice, query logs are commonly available and can provide partial, if not complete, set of true join relationships. We evaluate \sysname's result quality under varying percentages of available joins, which are randomly sampled from the ground truth. Figure~\ref{fig:quality_vs_sample_ratio} shows the F1 score of \sysname-EM as the proportion of known joins increases across the four datasets.

  Overall, increased availability of true joins via query logs leads to improved result quality for \sysname. This is because sampled true joins provide valuable structural signals regarding the connectivity of the underlying ground truth adjacency matrix. For instance, on TPC-DS with 40\% of true joins available, \sysname-GPT-4o achieves an F1 score of 0.91 and \sysname-XGBoost reaches an F1 score of 0.78. When the availability of true joins increases to 80\%, these scores rise to 0.96 and 0.92, respectively. The benefits of known joins are also significant on the REAL dataset, where F1 scores increase by over 20 points for both \sysname-XGBoost and \sysname-GPT-4o when 80\% of true joins are known. In contrast, the base models, XGBoost and GPT-4o, exhibit negligible improvement from increased known joins, as indicated by the flat dashed lines in Figure~\ref{fig:quality_vs_sample_ratio}. This is because base models already have (nearly) perfect recall on all datasets (see Table~\ref{tab:quality_comparison}). This gives strong evidence that the observed quality gains are driven specifically by our EM algorithm based on low-rank matrix completion.

\subsection{When Data Values are Available}\label{subsec:when_data_available}
  \begin{figure}[t]
    \centering
    \includegraphics[width=0.95\columnwidth]{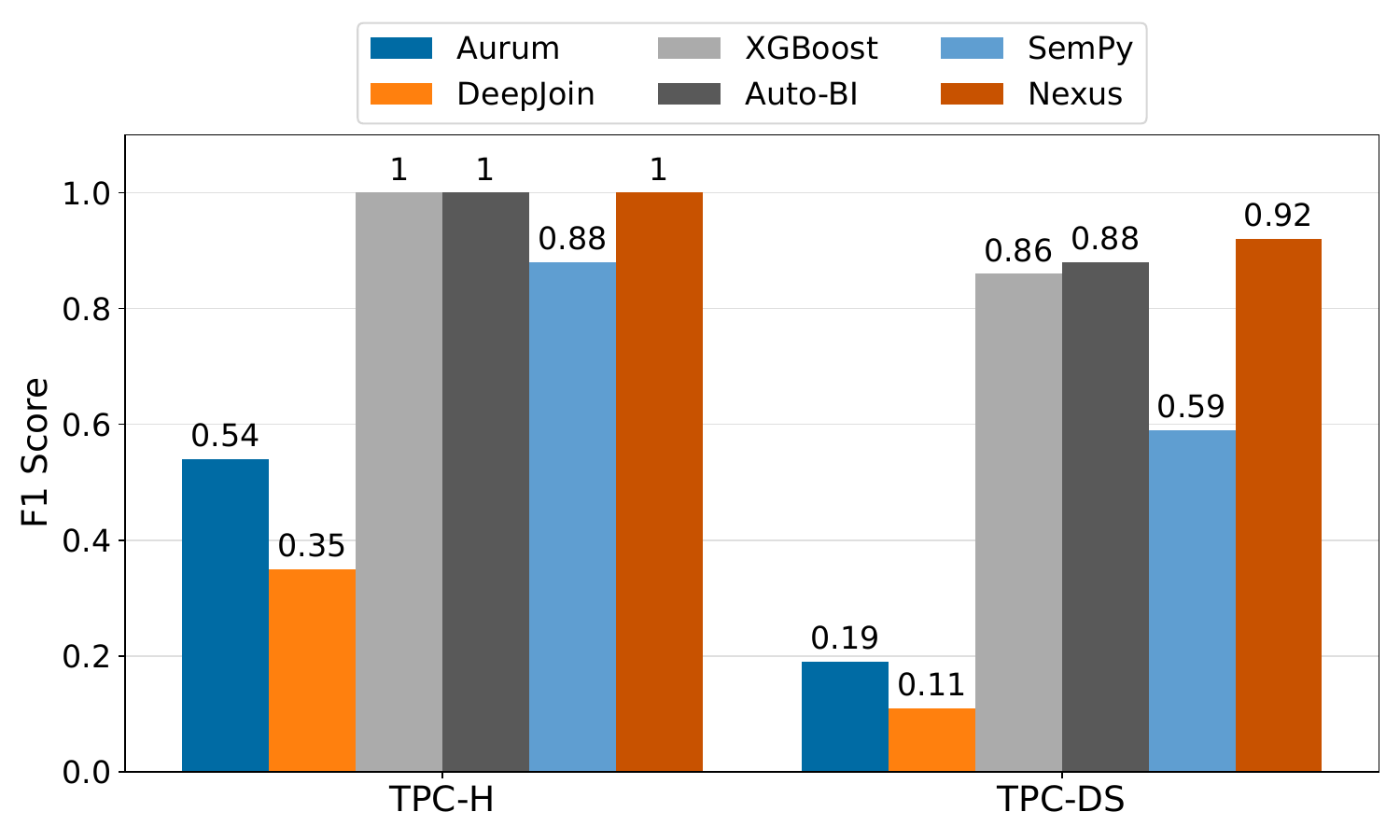}
    \caption{Comparison of F1 score on TPC-H and TPC-DS when data values are available.}
    \label{fig:f1_assuming_data_access}
  \end{figure}

  While \sysname is designed for enterprise settings where data access is restricted, it remains highly effective when data values are available. We compare \sysname-EM against Aurum, DeepJoin, XGBoost, Auto-BI, and SemPy on the TPC-H and TPC-DS datasets. The multi-source datasets are excluded in the interest of space. With full data access, we switch to Auto-BI's XGBoost model trained on additional features from data values across all relevant methods, replacing the metadata-only models in previous experiments.

  Figure~\ref{fig:f1_assuming_data_access} shows F1 scores for all six methods. Unlike the metadata-only setting, XGBoost, Auto-BI, and SemPy now perform competitively, highlighting their heavy reliance on data values. This contrast reinforces our earlier observation: when data signals are absent, utilizing the global join graph structure is critical for inferring joinability. Nevertheless, \sysname remains the top performer on both TPC-H and TPC-DS. Even though these datasets conform to the strict snowflake schema that Auto-BI is optimized for, \sysname still outperforms Auto-BI on TPC-DS.

  In addition to better ML predictions using value-based features, access to data values enables inclusion dependency tests, an effective technique for pruning join candidates. This test effectively eliminates many false positives that might otherwise appear highly joinable based on metadata alone. On TPC-DS, the number of join candidates drops from 3351 to 1232. On TPC-H, the reduction is even more dramatic, from 183 candidates to just 30. This aggressive pruning is a key reason why XGBoost, Auto-BI, and \sysname all achieve perfect F1 scores on TPC-H.

  Aurum and DeepJoin, representing approaches for the ``joinable table search" problem, trail behind all other methods. They require users to specify query columns and focused on a subset of true join keys in their original evaluation. In the setup of inferring full join graphs, both approaches produce a significant number of false positives when using all columns with unique values as query columns. DeepJoin further requires setting a fixed $k$ for top-$k$ search and is originally designed for string-typed columns, while over $80\%$ of joins in the test datasets involve numerical columns.

\section{RELATED WORK} \label{sec:related_work}
  Join detection has been studied extensively in the literature, in the context of PK/FK detection in relational databases~\cite{rostin2009machine, zhang2010multi, chen2014fast, jiang2020holistic}, BI model construction for analytical workloads~\cite{bharadwaj2021discovering, lin2023autobi}, and related table search in Web corpora~\cite{cafarella2008webtables, cafarella2009octopus, sarma2012finding, pimplikar2012answering} and data lakes~\cite{zhu2016lsh, fernandez2018aurum, zhu2019josie, bogatu2020d3l, zhang2020finding, dong2O21efficient, flores2021nextiajd, dong2O23deepjoin, deng2024lakebench, maynou2024freyja}. Prior work typically targets one of two scenarios: (1) joinable table search, a ``point query'' setup that finds tables joinable with a given input table/column, and (2) join graph inference, which finds all join relationships within a collection of tables. This work focuses on the latter without requiring user inputs.

  Traditional methods on join detection rely on heuristics like name similarity and value overlap~\cite{Microsoft/SemPy, chen2014fast}. For example, Chen et al.~\cite{chen2014fast} combine signals from table and column names with data values to distinguish true relationships from spurious candidates, using containment and other constraints to prune unlikely join candidates. Similarly, Jiang et al.~\cite{jiang2020holistic} jointly detect primary and foreign keys using a comprehensive set of pruning rules. However, these methods are often computationally expensive and miss semantic matches. Zhang et al.~\cite{zhang2010multi} detect PK/FK relationships by measuring distributional similarity between columns with the Earth Mover’s Distance, but this approach does not scale well, requiring over 2.5 hours to process 10 GB of data~\cite{chen2014fast}.

  Embedding-based approaches represent table and column metadata as vectors and use approximate nearest neighbor (ANN) search to find join candidates~\cite{bogatu2020d3l, dong2O23deepjoin, cong2023warpgate}. While these methods can uncover non-obvious semantic joins, their similarity metrics do not guarantee correctness, often leading to a high rate of false positives.

  Recent work has shifted toward ML-based methods. Rostin et al.~\cite{rostin2009machine} propose learning PK/FK relationships from known foreign keys using various ML models. However, subsequent work has shown that such pretrained models often fail to generalize to unseen datasets~\cite{chen2014fast, lin2023autobi}. In \sysname, we address this limitation, by adopting an EM algorithm that iteratively improve the predictions from the pretrained ML model with the guidance of LLMs.

  Our work is most closely related to recent approaches by \citet{bharadwaj2021discovering} and \citet{lin2023autobi}, which aim to automatically and efficiently construct complete join graphs with minimal user input. Similar to earlier work~\cite{rostin2009machine}, the method from~\citet{bharadwaj2021discovering} relies on ML models trained on large-scale data. While effective on in-distribution datasets, their approach does not address out-of-distribution generalization and overlooks the global structure of the join graph. Auto-BI~\cite{lin2023autobi} target a specific BI setting, where it constructs a join graph by applying a k-Min-Cost-Arborescence optimization on top of ML predictions. However, its enforcement of a snowflake schema is overly restrictive for datasets of loose schema. In contrast, our approach is more general and makes no assumptions about workload characteristics. We base our method on the key observations that join graphs in real-world scenarios are highly sparse and exhibit a low-rank structure. By leveraging this insight, we demonstrate that our approach performs robustly across a diverse range of datasets.

  Finally, most prior work, except~\cite{bharadwaj2021discovering}, assumes access to data values, but value-based feature computation on large tables is computationally expensive and often impractical in enterprise settings due to compliance and audit requirements~\cite{bharadwaj2021discovering}. In contrast, \sysname is designed to operate solely on metadata, while still providing high-quality results in an efficient manner.

\section{CONCLUSION} \label{sec:conclude}
  In this work, we introduce \sysname, a novel, metadata-only solution for join graph inference designed specifically for enterprise environments with stringent data privacy requirements. Our approach is founded on key insights derived from analyzing numerous real-world database schemas: join graphs are not only highly sparse but also consistently exhibit a low-rank structure. We exploit these properties by formulating join graph inference as a low-rank matrix completion problem. To overcome the poor generalization of existing pretrained models, we propose a novel EM algorithm that iteratively refines join probabilities by incorporating semantic compatibility assessments from LLMs. Extensive evaluations, including on a real production dataset, demonstrate that \sysname achieves significant improvements in both effectiveness and efficiency.

\bibliographystyle{ACM-Reference-Format}
\bibliography{sections/reference}

@String{Computing = "Computing" }

@String{Computer = "{IEEE} Computer" }

@String{Springer = "Springer-Verlag" }

@inproceedings{chistov1984lrmc,
  author       = {Alexander L. Chistov and
                  Dima Grigoriev},
  editor       = {Michal Chytil and
                  V{\'{a}}clav Koubek},
  title        = {Complexity of Quantifier Elimination in the Theory of Algebraically
                  Closed Fields},
  booktitle    = {Mathematical Foundations of Computer Science 1984, Praha, Czechoslovakia, September 3-7, 1984, Proceedings},
  series       = {Lecture Notes in Computer Science},
  volume       = {176},
  pages        = {17--31},
  publisher    = {Springer},
  year         = {1984},
}

@article{dempster1977maximum,
  title={Maximum likelihood from incomplete data via the EM algorithm},
  author={Dempster, Arthur P and Laird, Nan M and Rubin, Donald B},
  journal={Journal of the royal statistical society: series B (methodological)},
  volume={39},
  number={1},
  pages={1--22},
  year={1977},
  publisher={Wiley Online Library}
}

@article{cafarella2008webtables,
  author       = {Michael J. Cafarella and
                  Alon Y. Halevy and
                  Daisy Zhe Wang and
                  Eugene Wu and
                  Yang Zhang},
  title        = {WebTables: exploring the power of tables on the web},
  journal      = {Proc. {VLDB} Endow.},
  volume       = {1},
  number       = {1},
  pages        = {538--549},
  year         = {2008}
}

@article{candes2009lrmc,
  author       = {Emmanuel J. Cand{\`{e}}s and
                  Benjamin Recht},
  title        = {Exact Matrix Completion via Convex Optimization},
  journal      = {Found. Comput. Math.},
  volume       = {9},
  number       = {6},
  pages        = {717--772},
  year         = {2009}
}

@article{koren2009mftechniques,
  author       = {Yehuda Koren and
                  Robert M. Bell and
                  Chris Volinsky},
  title        = {Matrix Factorization Techniques for Recommender Systems},
  journal      = {Computer},
  volume       = {42},
  number       = {8},
  pages        = {30--37},
  year         = {2009}
}

@article{cafarella2009octopus,
  author       = {Michael J. Cafarella and
                  Alon Y. Halevy and
                  Nodira Khoussainova},
  title        = {Data Integration for the Relational Web},
  journal      = {Proc. {VLDB} Endow.},
  volume       = {2},
  number       = {1},
  pages        = {1090--1101},
  year         = {2009}
}

@inproceedings{rostin2009machine,
  author       = {Alexandra Rostin and
                  Oliver Albrecht and
                  Jana Bauckmann and
                  Felix Naumann and
                  Ulf Leser},
  title        = {A Machine Learning Approach to Foreign Key Discovery},
  booktitle    = {12th International Workshop on the Web and Databases, WebDB 2009,
                  Providence, Rhode Island, USA, June 28, 2009},
  year         = {2009},
}

@article{zhang2010multi,
  author       = {Meihui Zhang and
                  Marios Hadjieleftheriou and
                  Beng Chin Ooi and
                  Cecilia M. Procopiuc and
                  Divesh Srivastava},
  title        = {On Multi-Column Foreign Key Discovery},
  journal      = {Proc. {VLDB} Endow.},
  volume       = {3},
  number       = {1},
  pages        = {805--814},
  year         = {2010},
}

@inproceedings{sarma2012finding,
  author       = {Anish Das Sarma and
                  Lujun Fang and
                  Nitin Gupta and
                  Alon Y. Halevy and
                  Hongrae Lee and
                  Fei Wu and
                  Reynold Xin and
                  Cong Yu},
  title        = {Finding related tables},
  booktitle    = {Proceedings of the {ACM} {SIGMOD} International Conference on Management of Data, {SIGMOD} 2012, Scottsdale, AZ, USA, May 20-24, 2012},
  pages        = {817--828},
  publisher    = {{ACM}},
  year         = {2012}
}

@article{pimplikar2012answering,
  author       = {Rakesh Pimplikar and
                  Sunita Sarawagi},
  title        = {Answering Table Queries on the Web using Column Keywords},
  journal      = {Proc. {VLDB} Endow.},
  volume       = {5},
  number       = {10},
  pages        = {908--919},
  year         = {2012}
}

@inproceedings{jain2013minimization,
  author       = {Prateek Jain and
                  Praneeth Netrapalli and
                  Sujay Sanghavi},
  editor       = {Dan Boneh and
                  Tim Roughgarden and
                  Joan Feigenbaum},
  title        = {Low-rank matrix completion using alternating minimization},
  booktitle    = {Symposium on Theory of Computing Conference, STOC'13, Palo Alto, CA,
                  USA, June 1-4, 2013},
  pages        = {665--674},
  publisher    = {{ACM}},
  year         = {2013}
}

@article{chen2014fast,
  author       = {Zhimin Chen and
                  Vivek R. Narasayya and
                  Surajit Chaudhuri},
  title        = {Fast Foreign-Key Detection in Microsoft {SQL} Server PowerPivot for Excel},
  journal      = {Proc. {VLDB} Endow.},
  volume       = {7},
  number       = {13},
  pages        = {1417--1428},
  year         = {2014},
}

@article{zhu2016lsh,
  author       = {Erkang Zhu and
                  Fatemeh Nargesian and
                  Ken Q. Pu and
                  Ren{\'{e}}e J. Miller},
  title        = {{LSH} Ensemble: Internet-Scale Domain Search},
  journal      = {Proc. {VLDB} Endow.},
  volume       = {9},
  number       = {12},
  pages        = {1185--1196},
  year         = {2016}
}

@inproceedings{fernandez2018aurum,
  author       = {Raul Castro Fernandez and
                  Ziawasch Abedjan and
                  Famien Koko and
                  Gina Yuan and
                  Samuel Madden and
                  Michael Stonebraker},
  title        = {Aurum: {A} Data Discovery System},
  booktitle    = {34th {IEEE} International Conference on Data Engineering, {ICDE} 2018, Paris, France, April 16-19, 2018},
  pages        = {1001--1012},
  publisher    = {{IEEE} Computer Society},
  year         = {2018}
}

@article{kruse2018pyro,
  author       = {Sebastian Kruse and
                  Felix Naumann},
  title        = {Efficient Discovery of Approximate Dependencies},
  journal      = {Proc. {VLDB} Endow.},
  volume       = {11},
  number       = {7},
  pages        = {759--772},
  year         = {2018}
}

@inproceedings{zhu2019josie,
  author       = {Erkang Zhu and
                  Dong Deng and
                  Fatemeh Nargesian and
                  Ren{\'{e}}e J. Miller},
  title        = {{JOSIE:} Overlap Set Similarity Search for Finding Joinable Tables
                  in Data Lakes},
  booktitle    = {Proceedings of the 2019 International Conference on Management of
                  Data, {SIGMOD} Conference 2019, Amsterdam, The Netherlands, June 30
                  - July 5, 2019},
  pages        = {847--864},
  publisher    = {{ACM}},
  year         = {2019}
}

@inproceedings{zhang2020finding,
  author       = {Yi Zhang and
                  Zachary G. Ives},
  title        = {Finding Related Tables in Data Lakes for Interactive Data Science},
  booktitle    = {Proceedings of the 2020 International Conference on Management of Data, {SIGMOD} Conference 2020, online conference [Portland, OR, USA], June 14-19, 2020},
  pages        = {1951--1966},
  publisher    = {{ACM}},
  year         = {2020}
}

@article{jiang2020holistic,
  author       = {Lan Jiang and
                  Felix Naumann},
  title        = {Holistic primary key and foreign key detection},
  journal      = {J. Intell. Inf. Syst.},
  volume       = {54},
  number       = {3},
  pages        = {439--461},
  year         = {2020}
}

@inproceedings{bogatu2020d3l,
  author       = {Alex Bogatu and
                  Alvaro A. A. Fernandes and
                  Norman W. Paton and
                  Nikolaos Konstantinou},
  title        = {Dataset Discovery in Data Lakes},
  booktitle    = {36th {IEEE} International Conference on Data Engineering, {ICDE} 2020, Dallas, TX, USA, April 20-24, 2020},
  pages        = {709--720},
  publisher    = {{IEEE}},
  year         = {2020}
}

@article{bharadwaj2021discovering,
  author       = {Sagar Bharadwaj and
                  Praveen Gupta and
                  Ranjita Bhagwan and
                  Saikat Guha},
  title        = {Discovering Related Data At Scale},
  journal      = {Proc. {VLDB} Endow.},
  volume       = {14},
  number       = {8},
  pages        = {1392--1400},
  year         = {2021}
}

@inproceedings{dong2O21efficient,
  author       = {Yuyang Dong and
                  Kunihiro Takeoka and
                  Chuan Xiao and
                  Masafumi Oyamada},
  title        = {Efficient Joinable Table Discovery in Data Lakes: {A} High-Dimensional Similarity-Based Approach},
  booktitle    = {37th {IEEE} International Conference on Data Engineering, {ICDE} 2021, Chania, Greece, April 19-22, 2021},
  pages        = {456--467},
  publisher    = {{IEEE}},
  year         = {2021}
}

@inproceedings{flores2021nextiajd,
  author       = {Javier Flores and
                  Sergi Nadal and
                  Oscar Romero},
  title        = {Towards Scalable Data Discovery},
  booktitle    = {Proceedings of the 24th International Conference on Extending Database Technology, {EDBT} 2021, Nicosia, Cyprus, March 23 - 26, 2021},
  pages        = {433--438},
  publisher    = {OpenProceedings.org},
  year         = {2021}
}

@inproceedings{cong2023warpgate,
  author       = {Tianji Cong and
                  James Gale and
                  Jason Frantz and
                  H. V. Jagadish and
                  {\c{C}}agatay Demiralp},
  title        = {WarpGate: {A} Semantic Join Discovery System for Cloud Data Warehouses},
  booktitle    = {13th Conference on Innovative Data Systems Research, {CIDR} 2023,
                  Amsterdam, The Netherlands, January 8-11, 2023},
  publisher    = {www.cidrdb.org},
  year         = {2023}
}

@article{dong2O23deepjoin,
  author       = {Yuyang Dong and
                  Chuan Xiao and
                  Takuma Nozawa and
                  Masafumi Enomoto and
                  Masafumi Oyamada},
  title        = {DeepJoin: Joinable Table Discovery with Pre-trained Language Models},
  journal      = {Proc. {VLDB} Endow.},
  volume       = {16},
  number       = {10},
  pages        = {2458--2470},
  year         = {2023}
}

@article{kaminsky2023sawfish,
  author       = {Youri Kaminsky and
                  Eduardo H. M. Pena and
                  Felix Naumann},
  title        = {Discovering Similarity Inclusion Dependencies},
  journal      = {Proc. {ACM} Manag. Data},
  volume       = {1},
  number       = {1},
  pages        = {75:1--75:24},
  year         = {2023},
  url          = {https://doi.org/10.1145/3588929},
  doi          = {10.1145/3588929},
  bibsource    = {dblp computer science bibliography, https://dblp.org}
}

@article{lin2023autobi,
  author       = {Yiming Lin and
                  Yeye He and
                  Surajit Chaudhuri},
  title        = {Auto-BI: Automatically Build BI-Models Leveraging Local Join Prediction and Global Schema Graph},
  journal      = {Proc. {VLDB} Endow.},
  volume       = {16},
  number       = {10},
  pages        = {2578--2590},
  year         = {2023}
}

@inproceedings{li2024birdsql,
  author       = {Jinyang Li and
                  Binyuan Hui and
                  Ge Qu and
                  Jiaxi Yang and
                  Binhua Li and
                  Bowen Li and
                  Bailin Wang and
                  Bowen Qin and
                  Ruiying Geng and
                  Nan Huo and
                  Xuanhe Zhou and
                  Chenhao Ma and
                  Guoliang Li and
                  Kevin Chen{-}Chuan Chang and
                  Fei Huang and
                  Reynold Cheng and
                  Yongbin Li},
  title        = {Can {LLM} Already Serve as {A} Database Interface? {A} BIg Bench for
                  Large-Scale Database Grounded Text-to-SQLs},
  booktitle    = {Advances in Neural Information Processing Systems 36: Annual Conference on Neural Information Processing Systems 2023, NeurIPS 2023, New Orleans, LA, USA, December 10 - 16, 2023},
  year         = {2023}
}

@article{deng2024lakebench,
  author       = {Yuhao Deng and
                  Chengliang Chai and
                  Lei Cao and
                  Qin Yuan and
                  Siyuan Chen and
                  Yanrui Yu and
                  Zhaoze Sun and
                  Junyi Wang and
                  Jiajun Li and
                  Ziqi Cao and
                  Kaisen Jin and
                  Chi Zhang and
                  Yuqing Jiang and
                  Yuanfang Zhang and
                  Yuping Wang and
                  Ye Yuan and
                  Guoren Wang and
                  Nan Tang},
  title        = {LakeBench: {A} Benchmark for Discovering Joinable and Unionable Tables in Data Lakes},
  journal      = {Proc. {VLDB} Endow.},
  volume       = {17},
  number       = {8},
  pages        = {1925--1938},
  year         = {2024}
}

@article{pawitan2024llmconfidence,
  author       = {Yudi Pawitan and
                  Chris Holmes},
  title        = {Confidence in the Reasoning of Large Language Models},
  journal      = {CoRR},
  volume       = {abs/2412.15296},
  year         = {2024}
}

@inproceedings{mahaut2024factualconfidence,
  author       = {Mat{\'{e}}o Mahaut and
                  Laura Aina and
                  Paula Czarnowska and
                  Momchil Hardalov and
                  Thomas M{\"{u}}ller and
                  Llu{\'{\i}}s M{\`{a}}rquez},
  title        = {Factual Confidence of LLMs: on Reliability and Robustness of Current
                  Estimators},
  booktitle    = {Proceedings of the 62nd Annual Meeting of the Association for Computational Linguistics (Volume 1: Long Papers), {ACL} 2024, Bangkok, Thailand, August 11-16, 2024},
  pages        = {4554--4570},
  publisher    = {Association for Computational Linguistics},
  year         = {2024}
}

@article{cong2024openforge,
  author       = {Tianji Cong and
                  Fatemeh Nargesian and
                  Junjie Xing and
                  H. V. Jagadish},
  title        = {OpenForge: Probabilistic Metadata Integration},
  journal      = {Proc. {VLDB} Endow.},
  volume       = {18},
  number       = {9},
  pages        = {2914--2927},
  year         = {2025}
}

@online{Microsoft/SemPy,
  author = {Microsoft},
  title = {SemPy Python Library},
  url = {https://learn.microsoft.com/en-us/fabric/data-science/semantic-link-power-bi?tabs=sql},
  year = {2024},
  note = {Accessed: 2025-4-24}
}

@online{gpt-4o,
  author = {OpenAI},
  title = {GPT-4o},
  url = {https://openai.com/index/hello-gpt-4o/},
  organization = {OpenAI},
  year = {2024},
  note = {Accessed: 2025-4-24}
}

@article{dohmen2024schemapile,
  author       = {Till D{\"{o}}hmen and
                  Radu Geacu and
                  Madelon Hulsebos and
                  Sebastian Schelter},
  title        = {SchemaPile: {A} Large Collection of Relational Database Schemas},
  journal      = {Proc. {ACM} Manag. Data},
  volume       = {2},
  number       = {3},
  pages        = {172},
  year         = {2024}
}

@article{maynou2024freyja,
  author       = {Marc Maynou and
                  Sergi Nadal and
                  Raquel Panadero and
                  Javier Flores and
                  Oscar Romero and
                  Anna Queralt},
  title        = {{FREYJA:} Efficient Join Discovery in Data Lakes},
  journal      = {CoRR},
  volume       = {abs/2412.06637},
  year         = {2024}
}

\end{document}